\documentclass[a4paper,11pt]{article}
\usepackage{jheppub} 
\usepackage{lineno}
\usepackage{amsmath,amssymb,amsbsy,amstext,amsthm,booktabs,exscale,relsize,slashed,graphicx,amsfonts,upgreek,xcolor,amsmath,setspace}
\usepackage{multirow,dcolumn,bm,enumerate,url,hyperref}

\preprint{SLAC-PUB-17777}

\title{Dark Matter Raining on DUNE and Other Large Volume Detectors}

\author[a]{Javier F.~Acevedo}
\author[b]{Joshua Berger}
\author[c]{Peter B.~Denton}
\affiliation[a]{Particle Theory Group, SLAC National Accelerator Laboratory, Stanford, CA 94035, USA}
\affiliation[b]{Department of Physics, Colorado State University, Fort Collins, CO 80523, USA}
\affiliation[c]{High Energy Theory Group, Physics Department, Brookhaven National Laboratory, Upton, NY 11973, USA}

\emailAdd{jfacev@slac.stanford.edu} 
\emailAdd{joshua.berger@colostate.edu}
\emailAdd{pdenton@bnl.gov}

\abstract{Direct detection is a powerful means of searching for particle physics evidence of dark matter (DM) heavier than about a GeV with $\mathcal O(kiloton)$ volume, low-threshold detectors.
In many scenarios, some fraction of the DM may be boosted to large velocities enhancing and generally modifying possible detection signatures.
We investigate the scenario where 100\% of the DM is boosted at the Earth due to new attractive long-range forces. This leads to two main improvements in detection capabilities: 1) the large boost allows for detectable signatures of DM well below a GeV at large-volume neutrino detectors, such as DUNE, Super-K, Hyper-K, and JUNO, as possible DM detectors, and 2) the flux at the Earth's surface is enhanced by a focusing effect.
In addition, the model leads to a significant anisotropy in the signal with the DM flowing dominantly vertically at the Earth's surface instead of the typical approximately isotropic DM signal.
We develop the theory behind this model and also calculate realistic constraints using a detailed \texttt{GENIE} simulation of the signal inside detectors.}

\begin{document}
\maketitle
\flushbottom

\section{Introduction}
\label{sec:intro}
Dark matter (DM) direct detection \cite{Goodman:1984dc} is a powerful means of probing well motivated dark matter models at the GeV scale and heavier.
Lighter DM scenarios tend to be much harder to probe via scattering off a nucleus due to very low nuclear recoil thresholds.
Alternative detection approaches have been developed involving scattering off electrons, phonons, and other techniques, see $e.g.$ Refs.~\cite{Essig:2011nj,Essig:2022dfa}.
These can reach lower thresholds and can thus probe lighter DM, but detector volumes typically remain very small, at the tabletop size.

From the other perspective, it may well be that the nature of DM, and the associated dark sector in general, may lead to easier detection in some cases. This possibility has been extensively explored in boosted DM models, whereby some sub-component of halo DM acquires relativistic speeds, for example, through up-scattering interactions with cosmic-rays \cite{Bringmann:2018cvk,Alvey:2019zaa,Cappiello:2018hsu,Cappiello:2019qsw,Bell:2021xff,Maity:2022exk,Bell:2023sdq}, the Sun \cite{Kouvaris:2015nsa,An:2017ojc,Emken:2017hnp} or blazars \cite{Granelli:2022ysi}. Other possibilities include interactions with lighter dark sector states \cite{Berger:2014sqa,Agashe:2014yua,Giudice:2017zke,Geller:2022gey} as well as primordial black hole evaporation \cite{Calabrese:2021src}. In these scenarios, the detectability of DM is enhanced usually at the price of a decreased number density of easily detectable DM. 

Recently, in Ref.~\cite{Davoudiasl:2020ypv}, a model was developed where \emph{100\%} of the DM can be boosted at the Earth's surface due to the presence of a new long-range force (see also Ref.~\cite{Davoudiasl:2017pwe} where the same physical mechanism was used for the opposite effect). Along with any typical short-range force with regular matter, such a scenario may lead to a striking signature in a detector. In this paper, we investigate this possibility more generally and consider signatures in the largest volume detectors available: neutrino oscillation detectors. We explore in detail the different theoretical and phenomenological implications of a light scalar or vector mediator which boosts the DM by shifting its mass and energy, respectively. We find a number of interesting phenomena such as DM at the Earth's surface predominantly moving upward or downward -- a DM \textit{rain} type signal. On the detection side, we look at possible signatures in DUNE \cite{DUNE:2020ypp}, Super-K/Hyper-K \cite{Super-Kamiokande:2002weg,Hyper-Kamiokande:2018ofw}, IceCube/DeepCore \cite{IceCube:2011ucd}, and JUNO \cite{JUNO:2015sjr}, as well as the DM direct detection experiment LUX-ZEPLIN (LZ) \cite{LZ:2015kxe}. We calculate sensitivities using a specially modified version of \texttt{GENIE} \cite{Andreopoulos:2009zz,Andreopoulos:2015wxa} specifically suited to handle the nuclear physics signals of boosted DM \cite{Berger:2018urf}.

Our work is organized as follows: in Section~\ref{sec:BDM}, we discuss the class of models that can produce a DM rain. In Section~\ref{sec:boost_impac_b}, we compute the boost at the surface, as well as the maximum impact parameter for a DM particle to pass through a detector near the surface, as a function of the various model parameters. These quantities are used in Section~\ref{sec:det_prop} to compute the detection prospects of this effect using current and future large volume detectors. We conclude in Section~\ref{sec:outlook}. For the calculations throughout this paper, we assume a mostly negative metric signature and $\hbar = c = 1$.

\section{Boosted Dark Matter from Long-Range Forces}
\label{sec:BDM}
The DM rain can arise from new long-range interactions mediated by a light boson. With the correct signs on the couplings, this interaction produces an attractive potential, sourced by the SM matter or DM distribution within the Earth, raising in turn the escape velocity of incoming DM particles to relativistic speeds. This light boson may either couple directly to the Standard Model (SM), as previously investigated in Refs.~\cite{Davoudiasl:2017pwe,Davoudiasl:2020ypv}, or be confined within the dark sector, as recently explored in Ref.~\cite{Acevedo:2024zkg}. The latter scenario results in complex DM dynamics around the Earth because the potential is solely sourced by the DM captured by the Earth, which in turn affects the capture rate at future times. To illustrate the main features of our signal, we will focus on the case whereby the light boson also couples to the SM. In this case, the potential felt by the DM will be dominated by the static distribution of SM matter in the Earth. Depending on the specific model, captured DM might also contribute to the long-range potential, however we will conservatively neglect this effect. Furthermore, for the extremely small cross-sections detectable by large-volume detectors, the Earth's DM capture rate will be negligible for most of the parameter space.

\subsection{Models}
As our two benchmark models, we will consider a DM fermion $\chi$ coupled to either a light scalar $\phi$ or vector $A'_{\mu}$ with respective Lagrangians
\begin{equation}
    \mathcal{L} = i \bar{\chi} \slashed{\partial} \chi - m_\chi \bar{\chi} \chi + \frac{1}{2} (\partial \phi)^2 - \frac{1}{2} m_\phi^2 \phi^2 - g_\chi \phi \bar{\chi} \chi - g_{\rm SM} \phi \bar{\psi}_f \psi_f~,
    \label{eq:mod_sca}
\end{equation}
and 
\begin{equation}
    \mathcal{L} = i \bar{\chi} \slashed{\partial} \chi - m_\chi \bar{\chi} \chi - \frac{1}{4} F^\prime_{\mu\nu} F^{\prime \mu\nu} + \frac{1}{2} m_{A^\prime}^2 A^\prime_\mu A^{\prime \mu} - g_\chi A^\prime_\mu \bar{\chi} \gamma^\mu \chi - g_{\rm SM} A_\mu^\prime \bar{\psi}_f \gamma^\mu \psi_f~.
    \label{eq:mod_vec}
\end{equation}
Above, $\psi_f$ can be the SM nucleon or electron, and also couples to the new light boson with strength $g_{\rm SM} = g_n$ or $g_{\rm SM} = g_e$ respectively. In the long-range vector model, we additionally require the charge assignment to be such that the interaction between SM matter and DM is attractive. For simplicity, we have considered the case of fermionic DM, though there is no particular difficulty in translating our results to bosonic DM. Some example scalar couplings to nucleons can be realized through $e.g.$ a scalar coupling to the top quark or a colored vector-like generation \cite{Knapen:2017xzo}, or a CP-violating axion field $a$ with a non-derivative coupling of the form $\mathcal{L} \supset g_{aNN} a \bar{n} n$ \cite{Bigazzi:2019hav,Okawa:2021fto,Fan:2023hci}. Alternatively, one could also consider electrophilic couplings $\mathcal{L} \supset g_e \phi \bar{e} e$; see $e.g.$ Ref.~\cite{Parikh:2023qtk} for a specific model realization. For this scenario, existing limits on $g_e$ are weaker relative to nucleophilic couplings (see below). It is also worth noting that light scalars and pseudo-scalars have also been invoked in explanations for nuclear decay measurements \cite{Feng:2016ysn,Fornal:2017msy} and the neutron lifetime anomaly \cite{Fornal:2018eol}. On the other hand, long-range vectors have been considered in numerous SM extensions. Since we require the vector interaction to be net attractive, as an example one could consider U(1) theories with gauged baryon or lepton number, which have been explored in a number of works \cite{FileviezPerez:2010gw,Schwaller:2013hqa,Duerr:2013lka,Bell:2014tta,Tulin:2014tya,Batell:2014yra,Duerr:2017whl}.

To achieve a sizable boost, we require $m_{\phi , A^\prime }^{-1} \gg R_\oplus$, where $R_\oplus \simeq 6.37 \times 10^8 \ \rm cm \simeq 3.21 \times 10^{13} \ \rm eV^{-1}$ is the Earth's radius. On the other hand, in order to not disturb the DM distribution near the Earth, we will also limit our analysis to a range $m_{\phi,A^\prime}^{-1} \lesssim 1~\text{AU}$. A larger range may be possible, but then the force due to Sun on the DM will be larger than the force due to the Earth, which would modify the expected anisotropy in the signal. This is unlikely to run afoul of any astrophysical constraints on the DM distribution, but the resulting flux would differ somewhat from what we consider here; see Ref.~\cite{Davoudiasl:2023uiq} for some additional discussion. We will thus assume that the force range is $R_{\oplus} \ll m^{-1}_{\phi,A^{\prime}} \ll 1 \ \rm AU$ for the purposes of computing the terrestrial flux of DM. It is worth noting that in this scenario, the flux of boosted DM from up-scattering within the Sun, a process that has been extensively analyzed in the past \cite{Kouvaris:2015nsa,An:2017ojc,Emken:2017hnp}, could be severely suppressed due to the long-range force preventing the DM from evaporating \cite{Acevedo:2023owd}. So, while we also consider a short-range interaction (see below) that would permit such a process, its corresponding contribution to the flux at a terrestrial detector would be absent.

Within this picture, the Earth's distribution of SM matter will source a potential that accelerates incoming DM, increasing both the energy and impact parameters at which these particles could be detected. In both cases, the high density background within the Earth means $\langle \bar{\psi}_f \psi_f \rangle \simeq n_f$, where $n_f$ is the nucleon/electron density of the Earth given by the PREM model \cite{Dziewonski:1981xy}, so the scalar $\phi$ or time-like vector component $A'_0$ acquire a classical expectation value within and around the object. Similar scenarios involving long-range classical fields in celestial objects have recently been considered in various contexts, such as thermal particle production in stars \cite{DeRocco:2020xdt}, DM evaporation from stars and planets \cite{Acevedo:2023owd}, DM detection \cite{Davoudiasl:2017pwe,Davoudiasl:2020ypv}, dark kinetic heating of celestial bodies \cite{Gresham:2022biw,Acevedo:2024zkg}, Type-Ia supernova ignition and x-ray superbursts \cite{Raj:2023azx}, and matter effects in active and sterile neutrino oscillations \cite{Joshipura:2003jh,Berlin:2016woy,Brdar:2017kbt,Wise:2018rnb,Smirnov:2019cae,Davoudiasl:2023uiq}. Outside the source, we expect the field to decay with distance as   
\begin{equation}
\label{eq:potential_sca}
    \phi(r) = \phi(R_\oplus) \left(\frac{R_{\oplus}}{r}\right) \, e^{-m_\phi (r-R_{\oplus})}~,
\end{equation}
\begin{equation}
\label{eq:potential_vec}
    A^\prime_{\mu}(r) = A^\prime_\mu(R_\oplus) \, \left(\frac{R_{\oplus}}{r}\right) \, e^{-m_{A^\prime} (r-R_{\oplus})}~,
\end{equation}
where, in the absence of screening and in the regime where $m^{-1}_{\phi,A^\prime} \gtrsim R_\oplus$, the field at the boundary reads
\begin{equation}
    \phi(R_\oplus) =  - \frac{g_{\rm SM} N_\oplus}{4 \pi R_\oplus} \, e^{-m_\phi R_\oplus}\,,
\end{equation}
\begin{equation}
    A^\prime_\mu(R_\oplus) = - \delta_{\mu 0} \, \frac{g_{\rm SM} N_\oplus}{4 \pi R_\oplus} \, e^{-m_{A^\prime} R_\oplus}\,.
\end{equation}
Above, $N_\oplus \simeq 3.53 \times 10^{51}$ ($\simeq 1.76 \times 10^{51}$) is the total number of nucleons (electrons, assuming an average mass-to-charge ratio $A/Z \sim 2$) in the Earth depending on the specific coupling under consideration. 
Note that additional interactions that can be generically induced, such as a quartic term $\propto \phi^4$ in the scalar case modify the shape of the evolution of the field in radius \cite{Denton:2023iaa} and might suppress the boundary field values $\phi(R_\oplus)$ and $A^\prime_\mu(R_\oplus)$ above (although this can depend on the specific model realization, see $e.g.$ Ref.~\cite{Acevedo:2024zkg}).
From these fields, we define the scalar and vector potential under which DM particles are boosted as 
\begin{equation}
    \label{eq:potential_sca2}
    \Phi(r) = g_\chi \, \phi(r)~,
\end{equation}
\begin{equation}
    \label{eq:potential_vec2}
    V_{\mu}(r) = g_\chi \, A^\prime_{\rm \mu}(r)~. 
\end{equation}
The effective coupling strength of DM to this long-range potential is thus parameterized by the product $g_{\rm SM} \, g_\chi$. For the vector interaction, we have additionally verified that the resulting field gradient is well below the threshold for it to decay via pair-production (in analogy with the Schwinger effect) for the DM mass range considered. This is the case even accounting for the DM density enhancement that would occur around the Earth under such new attractive force, which is discussed further below.

In order to have a detectable signal in terrestrial experiments, the DM must have short range interactions with SM particles, in addition to the long range interactions described above.  For the purposes of this work, provided there are interactions with light quarks, there will be only small differences between the various choices of model. For concreteness, we consider a model with a heavy vector-like mediator, such that, in the low-energy effective theory, there are universal interactions of the form
\begin{equation}
    \mathcal{L} = \frac{1}{\Lambda^2} \, \sum_{f = q,e} \left(\overline{\psi}_f \, \gamma^\mu \, \psi_f\right) \, \left(\overline{\chi} \, \gamma_\mu \, \chi\right)~,
\end{equation}
where the sum is over the SM light quarks and electrons. Rather than parameterize this model in terms of the scale $\Lambda$, we will express our final sensitivity results in terms of a DM-proton interaction cross-section $\sigma_{\chi p}$.

\subsection{Existing Limits on Long-Range Forces}
There exist limits on the effective DM-SM coupling $g_{\rm SM} \, g_\chi$ due to fifth force searches as well as various tests of self-interacting dark matter. For the force range that we focus on, the strongest constraints on the SM coupling $g_{\rm SM}$ are set by the MICROSCOPE mission \cite{Berge:2017ovy,Fayet:2018cjy,MICROSCOPE:2022doy}. This experiment aims to observe violations of the weak equivalence principle. Such violations would appear if, for example, there are new light scalar interactions that do not universally couple to all forms of energy, leading to composition-dependent forces. Specifically, MICROSCOPE measures the differential acceleration between a platinum and titanium masses in orbit. For a fifth force that couples to baryon number, this has resulted in a constraint of order 
\begin{equation}
    g_{n} \lesssim 8 \times 10^{-25}~,
\end{equation}
although if additional effects such as screening are present \cite{Blinov:2018vgc} this bound may be loosened to those obtained by torsion balances \cite{Schlamminger:2007ht}. For proton-only or neutron-only couplings, the above can be rescaled by a factor $Z/A$ or $(A-Z)/A$ respectively, where $Z$ and $A$ is the atomic and mass number of the material used in the fifth force search. However, for simplicity, from hereon we will consider isospin-invariant couplings. For electrons, one can similarly rescale the above by a factor $m_n/m_e \simeq 1836$. On the DM side, Bullet Cluster and halo shape observations constrain the DM self-interaction cross-section to $1 - 10 \ \rm cm^2 \ g^{-1}$, which translates into a bound for the self-coupling $g_\chi$. For the light mediator regime we are interested in, whereby $m_{\phi,A^\prime} \ll m_\chi u$ with $u$ being the typical DM speed of the astrophysical system in question, the quantity of interest is the transfer cross-section (note that this will generally only differ by an $\mathcal{O}(1)$ factor from the viscosity cross-section regardless of whether the particles are distinguishable \cite{Tulin:2013teo}). This can be estimated from the point of closest approach based on the classical equations of motion \cite{Khrapak:2003kjw,Feng:2009hw}. Matching this cross-section onto the above yields an approximate limit for the coupling $g_\chi$, see $e.g.$ Refs.~\cite{Knapen:2017xzo,Coskuner:2018are,Gresham:2022biw}. For typical DM halo velocities, this is approximately
\begin{equation}
    g_\chi \lesssim 4 \times 10^{-6} \ \left(\frac{m_\chi}{\rm MeV}\right)^{3/4}~,
    \label{eq:bc_lim}
\end{equation}
if we require the self-interaction cross-section to be less than $\sim 1 \ \rm cm^2/g$. Note that, while light mediators can lead to large self-interaction cross-sections at dwarf galaxy scales, ongoing investigations have hinted at this possibility \cite{Ahn:2002vx,Agrawal:2016quu,Slone:2021nqd}. It is also worth noting these bounds can be significantly less robust compared to their SM counterpart due to their astrophysical nature~\cite{Robertson:2016xjh,Robertson:2022pjy,Popesso_2006, Wittman_2018,Adhikari:2022sbh}. They are also considerably weakened if one considers a DM sub-component. In any case, we take these bounds at face value in this work.

\section{Boost and Maximum Impact Parameter}
\label{sec:boost_impac_b}
The event rate of the DM rain will be determined by the flux of dark matter particles passing through a large-volume experiment. This in turn depends on the boost that the DM particles acquire from the long-range field, as well as the maximum impact parameter at which incoming DM particles can reach the Earth's surface. In what follows, we compute these quantities for the two benchmark long-range interactions detailed above. Notably, for the highly relativistic regime that is of interest in this work, the scalar and vector cases lead to qualitatively different fluxes, and so we will separately analyze them below. 

\subsection{Boost Factor at the Surface}
For a DM particle interacting with a background static scalar field $\Phi$ or vector field $V_{\mu}$, its Lagrangian is
\begin{equation}
    L = \begin{cases}
      \ - \sqrt{1 - v^2_\chi} \, (m_\chi + V_\mu U^\mu)  & \ \text{(vector)} \\
      \\
      \  - \sqrt{1 - v^2_\chi} \, (m_\chi + \Phi)  & \ \text{(scalar)} 
    \end{cases}\,,
\end{equation}
where $v_\chi$ is the DM's 3-velocity modulus and $U^\mu$ its 4-velocity. Note that, given the large boosts we focus on, we have completely neglected above the effect of the Earth's gravitational field. This results in the following equations of motion,
\begin{equation}
\label{eq:eom_main}
    \begin{cases}
      \ \dfrac{dp^{\mu}}{d\tau} = \dfrac{d}{d\tau} \left( m_\chi U^\mu + V^\mu \right) = U^\nu \partial^\mu V_\nu  & \ \text{(vector)} \\
      \\
      \ \dfrac{dp^{\mu}}{d\tau} = \dfrac{d}{d\tau} \left((m_\chi +\Phi) U^\mu\right) = \partial^{\mu} \Phi & \ \text{(scalar)}
    \end{cases}~,
\end{equation}
where $p^{\mu}$ is the 4-momentum in the Earth's rest frame and $\tau$ is the particle's proper time. We are interested in the velocity as a function of position, which can be easily obtained in both cases by taking the $\mu = 0$ component and integrating, yielding the DM's energy as a constant of motion,
\begin{equation}
\label{eq:eom_E}
    E = {\rm const.} = \begin{cases}
      \ \frac{m_\chi}{\sqrt{1-v_\chi^2}} + V_0 & \ \text{(vector)} \\
      \\
      \  \frac{m_\chi + \Phi}{\sqrt{1-v_\chi^2}} & \ \text{(scalar)}
    \end{cases}\,.
\end{equation} 
The energy is fixed by the initial condition. For the purposes of computing the final boost achieved at the Earth's surface, it suffices to neglect the initial halo kinetic energy of the DM, which is insignificant compared to the boosts of interest. We thus fix the energy here as $E \simeq m_\chi$. Note that the remaining components of Eq.~\eqref{eq:eom_main} yield the trajectory, $i.e.$ the time dependence of the coordinate $r = r(\tau)$. However, we are only interested in the boost that is reached at the surface, which we can simply obtain by fixing the energy above and solve for $v_\chi$ as a function of $r$. For each interaction benchmark, this yields
\begin{equation}
\label{eq:boost_main}
    v_\chi(r) = \begin{cases}
      \ \sqrt{\frac{\left(1 - \frac{V_0(r)}{m_\chi}\right)^2-1} {\left(1 - \frac{V_0(r)}{m_\chi}\right)^2}}  & \ \text{(vector)} \\
      \\
      \  \sqrt{1 - \left(1 + \frac{\Phi(r)}{m_\chi}\right)^2} & \ \text{(scalar)}
    \end{cases}\,.
\end{equation} 

In the scalar case, the DM velocity seemingly reaches the speed of light at the radius $r$ where $m_\chi + \Phi(r) = 0$, or in other words, when the effective, spacetime-dependent mass induced by the scalar interaction vanishes. 
However, once a sufficiently large boost is achieved, radiation losses and other quantum effects become significant, which prevent this from occurring. We will only consider the parameter space where this regime is not reached at the Earth's surface, guaranteeing at least the downward component of the DM rain. The upward component may be suppressed depending on the extent to which this radiative regime is reached within the Earth's interior; this is left for future investigation. To be conservative, in our sensitivity projections for the scalar force scenario, we rescale the flux by a factor $\sim 1/2$, accounting only for the downward DM rain flux. In contrast with the scalar case, only in the limit that the vector potential $|V_0| \rightarrow \infty$ the particle seemingly reaches the speed of light. In the vector case, we will consider both the upward and downward components of the flux. As before, at sufficiently large boosts radiative losses also become significant in this case. We analyze energy losses due to radiation in App.~\ref{app:rad_thresh}, but we note here that the boost at which these become significant lies considerably beyond what we simulate here. 

Additionally, for the vector case, we must consider the effect that the transiting DM has on the potential sourced by the Earth, as these particles, in order to be boosted, must necessarily be oppositely-charged. This can weaken the potential, as the Earth's net charge would be effectively reduced. As we show below, the flux of DM particles passing through the Earth is enhanced by a factor $\gamma^2$. Using kinetic theory arguments, it can be shown that this consequently leads to an enhancement of the Earth's internal DM density by a factor $\sim \gamma^2$ relative to the halo density. Note that this is different from gravitational capture, which requires the DM particles to scatter and lose energy; in this case, the density of \textit{unbound} particles is enhanced. Comparing the DM-sourced potential to the Earth's potential ($cf.$ Eq.~\eqref{eq:potential_vec2}), we find this correction to be negligible in most of the parameter space for self-couplings $g_\chi$ below the limit Eq.~\eqref{eq:bc_lim}.

\begin{figure}[htbp]
\centering
\includegraphics[width=.47\textwidth]{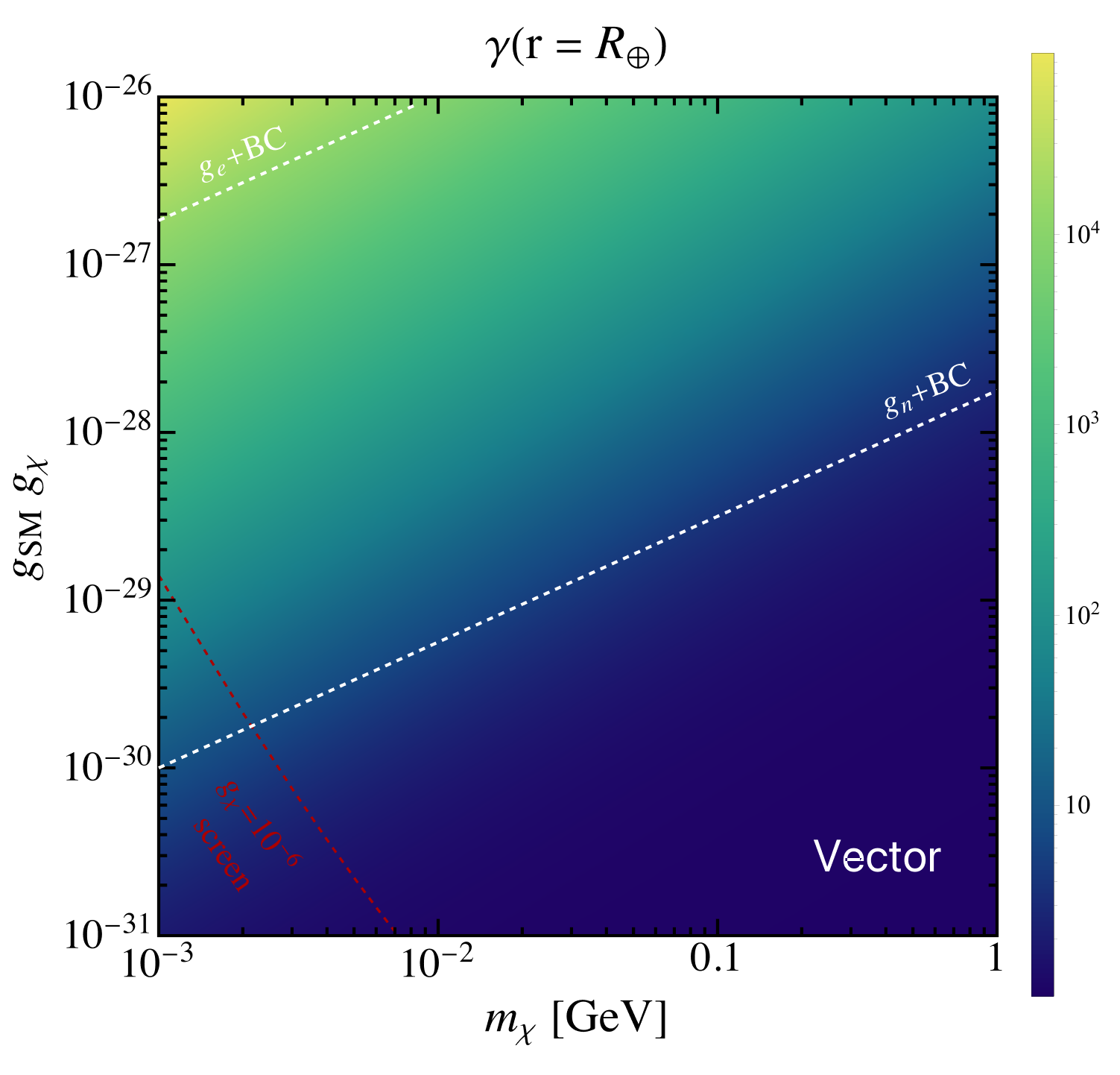}
\qquad
\includegraphics[width=.47\textwidth]{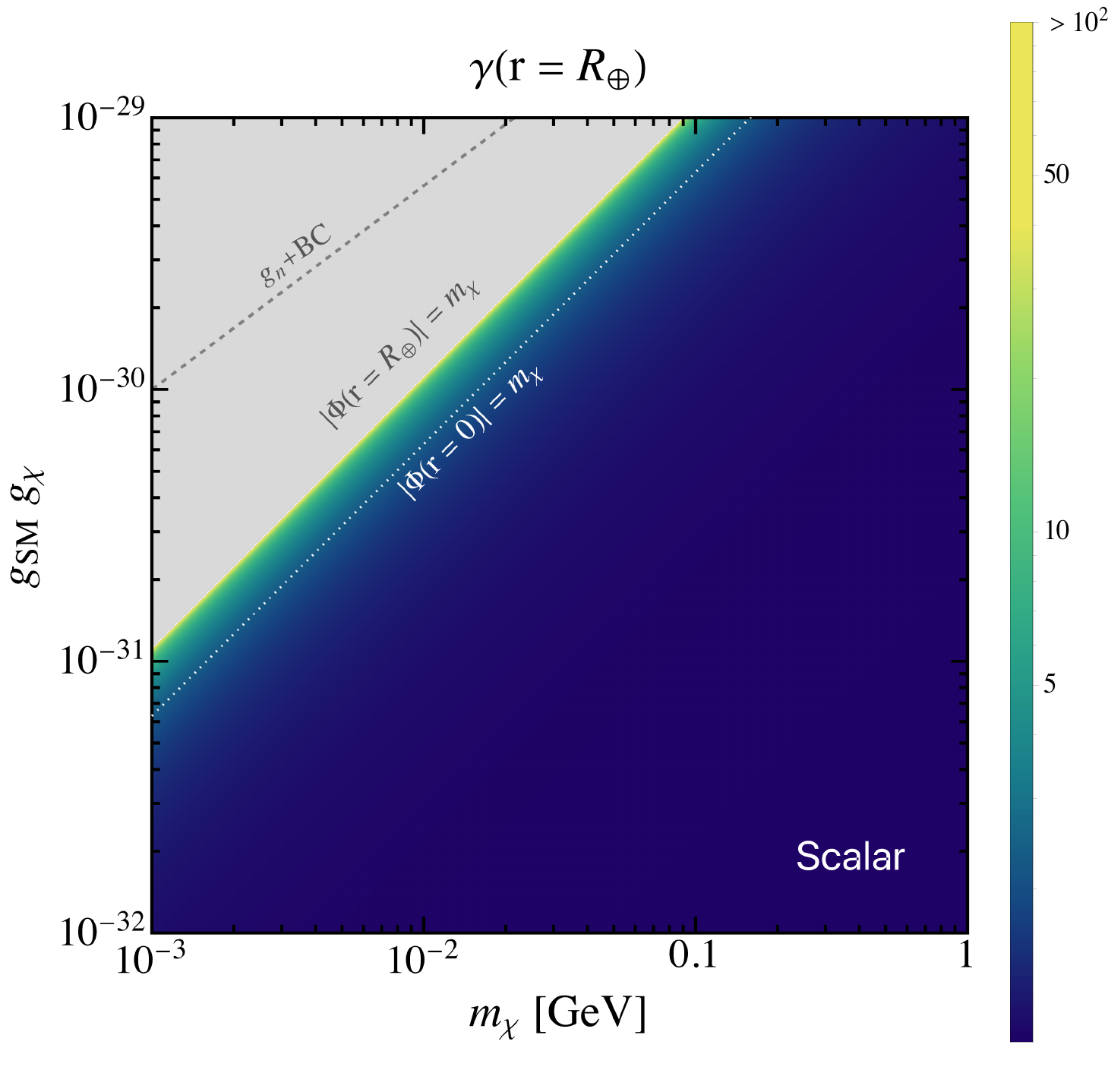}
\caption{\textbf{Left:} Boost factor at the Earth's surface, as a function of coupling strength and DM mass, for a vector long-range interaction.  We also indicate the maximum coupling allowed by electro- and nucleo-philic fifth force searches combined with DM self-interaction constraints from observations of the Bullet Cluster (\textit{white dashed}). The DM density enhancement within the Earth due to the attractive force can lead to screening in the vector case; we show this for a fiducial coupling of $g_\chi = 10^{-6}$ (\textit{red dashed}).
\textbf{Right:} Same as left panel, but for a long-range scalar interaction. 
In this case, we also show the parameter space where DM radiative effects are relevant at the Earth's center (\textit{white dotted}) and at the surface (\textit{gray shaded}).}
\label{fig:boost}
\end{figure}

Figure~\ref{fig:boost} shows the resulting DM boost at the Earth's surface, as a function of the coupling product $g_{\rm SM} \, g_\chi$ and DM mass $m_\chi$ for both benchmark interaction types. These are calculated in the limit $m^{-1}_\phi \gg R_{\oplus}$. For reference, we have included the combined limit from fifth force searches and Bullet Cluster dynamics on the coupling as a function of DM mass, for electron and nucleon couplings. In the scalar case, we also show the parameter space for which the radiative effects detailed above become significant at the Earth's center and surface, respectively indicated by the dashed line and shaded area. The parameter space where the vector potential may be weakened by transiting DM, as described above, is also indicated for the vector case for a fiducial coupling $g_\chi = 10^{-6}$. Overall, it can be seen that, for both interaction benchmarks, a wide range of boosts can be achieved. However, for a long-range scalar, these occur in a narrower region of parameter space due to the scalar backreaction effects becoming important for much lower couplings compared to the vector case.

\subsection{Maximum Impact Parameter}
We now compute the maximum impact parameter that results in a DM particle reaching the Earth's surface. As before, we will ignore the effects of the Earth's gravitational field, since we consider a net attractive long-range force with a coupling strength that is large relative to gravity. We will work in the Hamilton-Jacobi formalism (see $e.g.$ Ref.~\cite{landau2013classical}), which allows us to compute this quantity rather directly in terms of the particle's initial energy. The Hamilton-Jacobi equation for each benchmark interaction reads
\begin{equation}\label{eq:HJ-1}
    \begin{cases}
      \ \left(-\dfrac{\partial S}{\partial x^\mu} + V_\mu \right)\left(-\dfrac{\partial S}{\partial x_\mu} + V^\mu \right) = m_\chi^2 & \ \text{(vector)} \\
      \\
      \  \dfrac{\partial S}{\partial x^\mu} \dfrac{\partial S}{\partial x_\mu} = (m_\chi + \Phi)^2 & \ \text{(scalar)} \\
    \end{cases}
\end{equation}
where above $x = (t,r,\varphi)$\footnote{We assume without loss of generality that the orbit is contained in a plane, so only one azimuthal angle $\varphi$ is needed}, and $S(x)$ is the action of the DM particle evaluated along its physical trajectory. Eq.~\eqref{eq:HJ-1} is a differential equation for $S(x)$, which can solved through the ansatz 
\begin{equation}
    S(x) = E \, t - L \, \varphi + S_r(r) + \rm const.~.
    \label{eq:HJeq_sol}
\end{equation}
Above, $L$ is the DM's orbital angular momentum. The energy $E$ has been already computed in Eq.~\eqref{eq:eom_E}. We express the angular momentum as
\begin{equation} 
   \label{eq:orb_ang_mom}
    L = b \, m_\chi \, u~,
\end{equation}
where $b$ is the impact parameter, and $u$ is the DM's velocity far from the Earth where the effect of both gravity and the new long-range force are negligible. Lastly, the function $S_r(r)$ exclusively contains the radial dependence, and its derivative, is related to the radial velocity via 
\begin{equation}
    \frac{\partial S_r}{\partial r} = \begin{cases}
      \ - m_\chi \, \dfrac{dr}{d\tau} & \ \text{(vector)} \\
      \\
      \ - (m_\chi + \Phi) \, \dfrac{dr}{d\tau} & \ \text{(scalar)} \\   
      \end{cases}
\end{equation}
Replacing the derivatives of the action $S$ into Eq.~\eqref{eq:HJ-1} yields a direct relation between the radial velocity at a given point and the angular momentum $L$, which reads
\begin{equation}
   \label{eq:HJ-2}
   \left(\frac{dr}{d\tau}\right)^2 = \begin{cases}
      \ \left(\dfrac{E - V_0}{m_\chi}\right)^2 - \left(\dfrac{L}{m_\chi r}\right)^2 - 1 & \ \text{(vector)} \\
      \\
      \ \left(\dfrac{E}{m_\chi + \Phi}\right)^2 - \left(\dfrac{L}{(m_\chi + \Phi) r}\right)^2 - 1 & \ \text{(scalar)} \\   
      \end{cases}
\end{equation}

The maximum impact parameter will be defined by the maximum angular momentum $L$ at which a particle passes through a detector near $r \simeq R_{\oplus}$ with no radial velocity, $i.e.$ $dr/d\tau = 0$ at that point. This yields 
\begin{equation}\label{eq:max-impact}
    b_{\rm max} = \begin{cases}
      \ \dfrac{R_{\oplus}}{u} \sqrt{\left(\dfrac{E - V_0(R_{\oplus})}{m_\chi}\right)^2 - 1} \, = \, R_{\oplus} \, \gamma \left(\dfrac{v_\chi}{u}\right)  & \ \text{(vector)} \\
      \\
      \ \dfrac{R_{\oplus} \left(m_\chi + \Phi(R_{\oplus})\right)}{m_\chi u} \sqrt{\left(\dfrac{E}{m_\chi + \Phi(R_{\oplus})}\right)^2 - 1} \, = \, R_{\oplus} \left(\dfrac{v_\chi}{u}\right) & \ \text{(scalar)} \\
    \end{cases}
\end{equation}
In the rightmost expression for each case, we have used Eq.~\eqref{eq:eom_E}, and we have abbreviated $v_\chi = v_\chi(r = R_\oplus)$ and similarly for the boost factor $\gamma$. The maximum angular momentum for which a DM particle reaches a detector close to the Earth's surface, in terms of the boost achieved, is thus
\begin{equation}
    L_{\rm max} = b_{\rm max} \, m_\chi \, u \simeq \begin{cases}
      \ R_{\oplus} \, m_\chi \, \gamma \, v_\chi & \ \text{(vector)} \\
      \\
      \  R_{\oplus} \, m_\chi \, v_\chi & \ \text{(scalar)} \\
    \end{cases}
    \label{eq:J_max}
\end{equation}
where it is worth noting that the long-range vector enhances the impact parameter by a factor $\gamma$. As we show below, this results in significant differences in the sensitivity projections between the two benchmarks in the highly boosted regime. 

While the above estimates are valid for a force of any range, depending on the exact values of the coupling strength and mediator mass, the competition between the centrifugal term proportional to $(L/r)^2$ and the exponentially-decaying potential may induce a secondary centrifugal barrier at distances $r \gtrsim m^{-1}_{\phi,A^{\prime}}$. This can result in an effective impact parameter for reaching a detector that is substantially smaller than what Eq.~\eqref{eq:max-impact} predicts, consequently suppressing the flux. However, we have verified that, for the local DM halo speed and the range of boosts we have simulated, this barrier is overcome provided that $m^{-1}_{A^{\prime}} \gtrsim 100 \, R_{\oplus}$ in the vector case, and $m^{-1}_{\phi} \gtrsim 10 \, R_{\oplus}$ in the scalar case. Additional details of this effect are presented in App.~\ref{sec:centrifugal}. Even in the scenario where this secondary barrier is important, optimistic detection prospects would still be recovered. For simplicity, we will assume that the mediator is sufficiently long-ranged so that Eq.~\eqref{eq:max-impact} is in fact the maximum impact parameter at which a DM particle passes through a detector near the surface.

\begin{figure}[htbp]
\centering
\includegraphics[width=.47\textwidth]{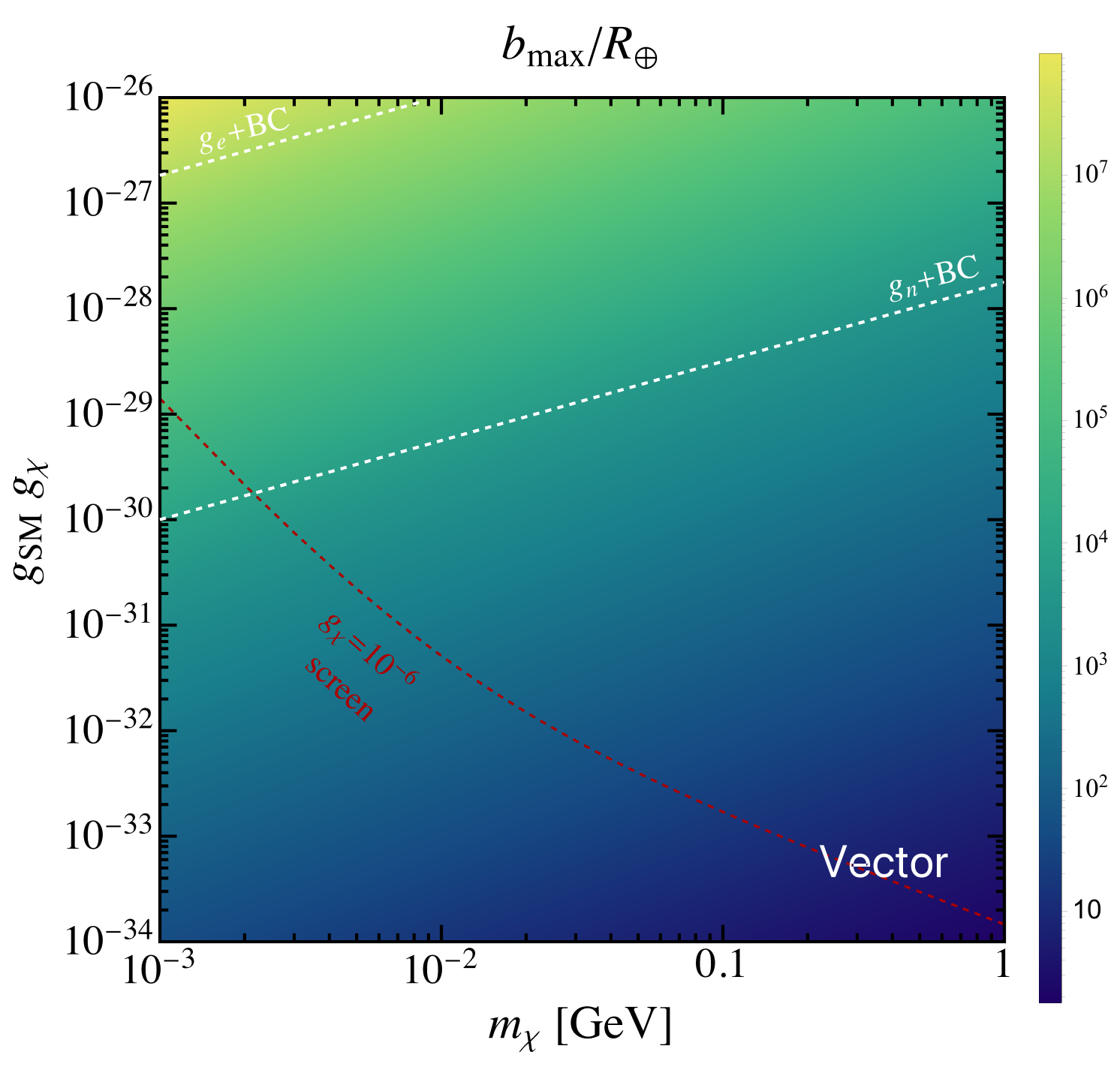}
\qquad
\includegraphics[width=.47\textwidth]{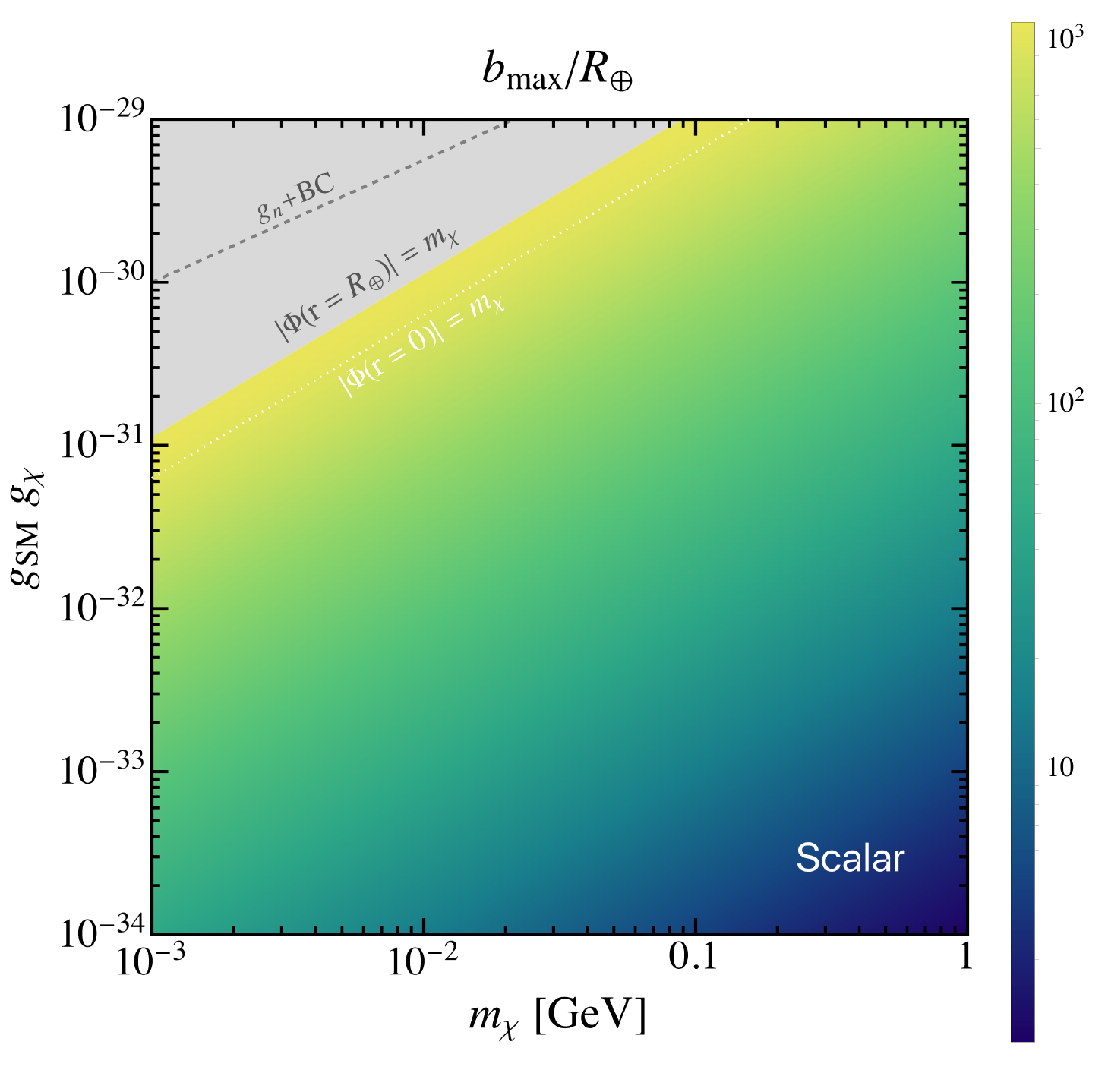}
\caption{Same as Fig.~\ref{fig:boost}, but for the maximum impact parameter $b_{\rm max}$, normalized to the Earth's radius, for which a DM particle passes through a detector. This assumes the mediator is sufficiently light, see main text.}
\label{fig:bmax}
\end{figure}

Figure~\ref{fig:bmax} shows the maximum impact parameter $b_{\rm max}$ for reaching a detector near the surface as a function of dark matter mass $m_\chi$ and coupling $g_{\rm SM} \, g_{\chi}$. To illustrate clearly the focusing effect of the long-range force, we have normalized $b_{\rm max}$ to the Earth radius $R_{\oplus}$, as this quantity is approximately what $b_{\rm max}$ would be if the Earth's gravitational field was the only long-range force acting on the DM. The difference between the scalar and vector cases is also visible: for the scalar interaction, already at moderate boosts we have $v_{\chi} \simeq 1$ and the impact parameter $b_{\rm max}$ saturates to a maximum determined by $u \simeq 10^{-3}$ for halo DM. By contrast, in the vector case $b_{\rm max}$ is enhanced by a factor $\gamma \gg 1$.  

\section{Detection Prospects}
\label{sec:det_prop}
With the above determinations of the boost and maximum impact parameter for DM by new attractive long-range fields, we now determine its rate of interaction in various large volume detectors.

\subsection{Flux through Detector}
The differential flux of DM particles passing through is given by \cite{Gould:1987ir}
\begin{equation}
d\mathcal{F} = \pi \, \frac{f(u)}{u} \, du \, dJ^2~,
\label{eq:flux_main}
\end{equation}
where $J = L/m_\chi$ is the conserved angular momentum per unit rest mass ($cf.$ Eq.~\eqref{eq:orb_ang_mom}), $u$ is the DM velocity far from the Earth, and $f(u)$ is the velocity distribution. For ease of reference, we provide additional details on how Eq.~\eqref{eq:flux_main} is derived in App.~\ref{sec:Gould_recap}. We will find that only certain averages taken over the DM distribution are relevant. When necessary, we take these averages using the distribution $f(u)$ obtained by Ref.~\cite{Necib:2018iwb}. Note that we take the normalization so that
\begin{equation}
    \int_0^\infty f(u) \, du = n_\chi~.
\end{equation}
Above, $n_\chi = \rho_\chi/m_\chi$ is the local DM number density. For the DM mass density, we fix $\rho_\chi = 0.4 \ \rm GeV \ cm^{-3}$ \cite{Pato:2015dua,2024MNRAS.528..693O}. 

While we are mostly interested in the differential flux above to compute event rates, for completeness, we also provide the integrated DM flux passing through a detector, which reads
\begin{equation}
\label{eq:flux-tot}
    \frac{d\mathcal{F}}{dA} = \frac{1}{4\pi R^2_{\oplus}} \times \left(\pi J^2_{\rm max} \left\langle \frac{1}{u} \right\rangle\right) \simeq \frac{1}{4} \, n_\chi \, v^2_\chi \, \left\langle \frac{1}{u}\right\rangle \times
    \begin{cases}
      \ \gamma^2 & \ \text{(vector)} \\
      \\
      \ 1/2 & \ \text{(scalar)} \\
    \end{cases}~,
\end{equation}
where we have defined the average
\begin{equation}
\left\langle g(u) \right\rangle = \frac{1}{n_\chi} \int_0^\infty  g(u) f(u) \, du~.
\end{equation} 
and the extra factor of $1/2$ in the scalar case is a penalty factor from DM particles that may enter a radiative regime while passing through the Earth, $cf.$ Sec.~\ref{sec:boost_impac_b}. In particular, the average inverse speed can be approximated as
\begin{equation}
\left\langle \frac{1}{u}\right\rangle \simeq \sqrt{\frac{2}{\pi}} \frac{1}{\sigma_u}~,
\label{eq:mean 1/u}
\end{equation}
where $\sigma_u \simeq 10^{-3}$ is the velocity dispersion of the DM at the local position.  While Eq.~\eqref{eq:mean 1/u} is a good approximation for the mean inverse velocity, we calculate the average here using a recent distribution derived from the Gaia survey~\cite{Necib:2018iwb}. Note that the maximum angular momentum per rest mass $J_{\rm max}$, along with the DM's contact cross-section and mass, will determine the event rate in a detector. 

\subsection{Event Rate}
First, let us consider a small volume element located at a distance $r$ from the center of the Earth. A DM particle entering this volume at angle $\theta$ with respect to the radial direction will travel a distance 
\begin{equation}
d\ell = \frac{d\ell}{dr} \, dr = \frac{d(r/\cos\theta)}{dr} \, dr = \frac{dr}{\cos\theta}~,
\end{equation}
where $dr$ is the radial extent of the volume element. We relate this angle $\theta$ to the conserved angular momentum using $J \propto \sin\theta$, which yields 
\begin{equation}
\label{eq:cos_theta}
\cos\theta = \sqrt{1 - \left(\frac{J}{J_{\rm max}}\right)^2}~.
\end{equation}
For convenience, we have parameterized the angle $\theta$ in terms of the maximum angular momentum $J_{\rm max} = L_{\rm max}/m_\chi$ for which a DM particle reaches the a detector near the surface with $\theta = \pi/2$, $cf.$ Eq.~\eqref{eq:J_max}. 

The probability for a dark matter particle undergoing this trajectory to interact is
\begin{equation}
dP = \sum_i \sigma_{\chi i} \, n_i \, d\ell = \frac{1}{\sqrt{1 - \left(J/J_{\rm max}\right)^2}} \, \sum_i \sigma_{\chi i} \, n_i \,dr ~,
\end{equation}
where $n_i$ is the number density of target nuclei $i$ and $\sigma_{\chi i}$ is the cross section for interaction with target nucleus $i$. As necessary, one should sum over the various isotopes making up the detector.  The differential interaction rate $d\Gamma$ in a small volume element at $r = R_\oplus$ will be given by the product between flux, the fraction of total area $4 \pi R_\oplus^2$ spanned by the volume element, and the probability of interaction,
\begin{equation}
d\Gamma = 2 \, dP \, d\mathcal{F} \, \frac{dA}{4 \pi R_\oplus^2} ~. 
\end{equation}
Above, we have included an additional factor of 2, as the DM can pass through this volume in either direction. Using Eq.~\eqref{eq:flux_main}, we obtain
\begin{equation}
d\Gamma = \frac{f(u)}{2u} \frac{\sum_i \sigma_{\chi i} \, n_i}{\sqrt{1 - \left(J/J_{\rm max}\right)^2}} \, \frac{dV}{R_\oplus^2} \, du \, dJ^2~. 
\label{eq:diff_rate_1}
\end{equation}
where we have defined $dV = dA \, dr$. The integration over angular momentum $J$ can be trivially performed, and yields a factor $2 J^2_{\rm max}$. Further integration over the DM halo velocity $u$ yields the differential interaction rate per unit volume
\begin{equation}
\frac{d\Gamma}{dV} = \sum_i n_i \, n_\chi \, \left\langle \sigma_{\chi i} \, \frac{J^2_{\rm max}}{u  R_\oplus^2}\right\rangle~. 
\label{eq:diff_rate_2}
\end{equation}

Using Eq.~\eqref{eq:J_max} and assuming a homogeneous detector, we can write the total number of events expected during its livetime $T$ in terms of the final boost at the Earth's surface $w$ as 
\begin{equation}
    N_{\rm event} = T \, n_\chi \, \sum_i \, N_i \, \sigma_{\chi i} \, v_\chi^2 \, \left\langle \frac{1}{u}\right\rangle \times \begin{cases}
      \  \gamma^2 & \ \text{(vector)} \\
      \\
      \ 1/2 & \ \text{(scalar)} \\
    \end{cases}
\end{equation}
where integrating Eq.~\eqref{eq:diff_rate_2} over volume simply yields the total number $N_i$ of nuclei $i$ in the detector. We have also factorized $J_{\rm max}$, since this depends on $v_\chi$, which is nearly independent of $u$ in the highly boosted limit of interest here ($cf.$ Eq.~\eqref{eq:boost_main}), as well as the cross-section.
Relative to the scalar interaction case, the rate for a vector long-range force is enhanced by a factor of $\gamma^2$, which may be quite large in the regime we focus on. On the other hand, we have explicitly included a factor $1/2$ in the scalar case, to conservatively account for possible radiative effects while passing through the Earth, which may suppress the up-going component of the flux depending on the parameter space. 

\subsection{Projected Reach of Large Volume Detectors}
\begin{figure}[!tbh]
\centering
\includegraphics[width=0.45\textwidth]{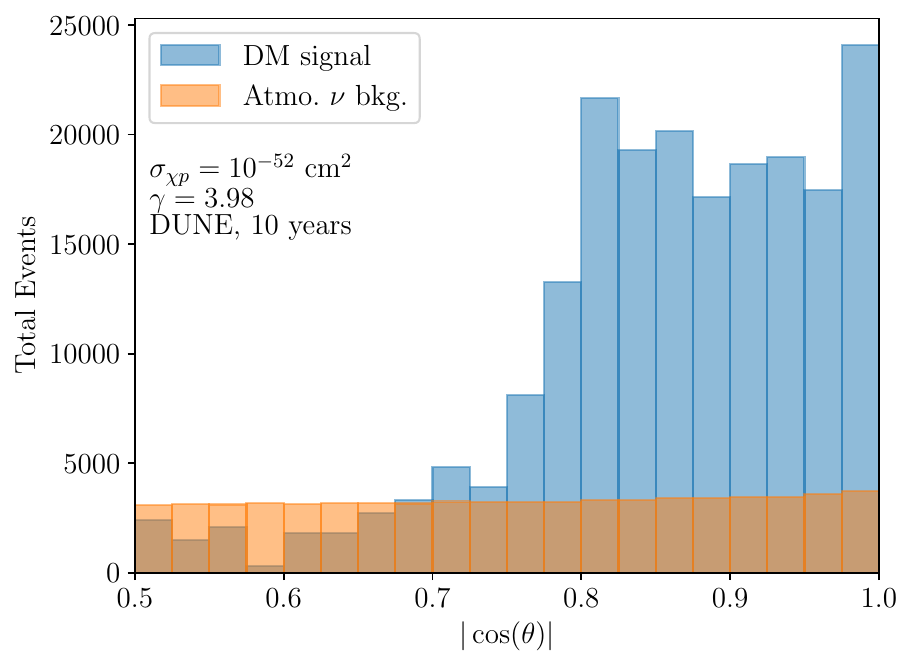}
\includegraphics[width=0.45\textwidth]{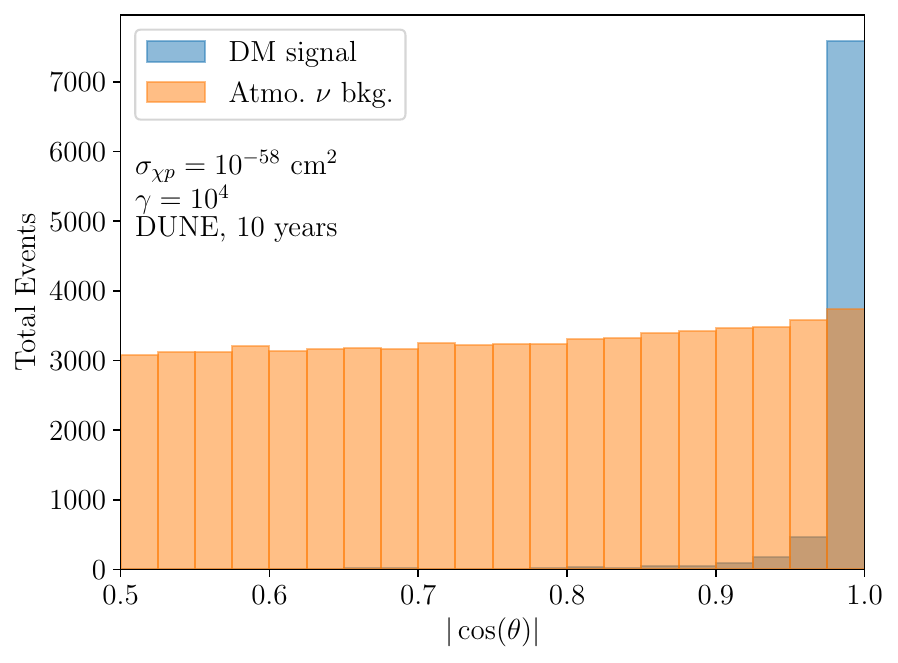}
\caption{Distribution of the cosine of the angle with respect to the vertical at a boost of $\gamma = 3.98$ (left) and $\gamma = 10^4$ (right) and a DM mass of $100~\text{MeV}$.}\label{fig:angle}
\end{figure}

\begin{figure}[!tbh]
\centering
\includegraphics[width=0.45\textwidth]{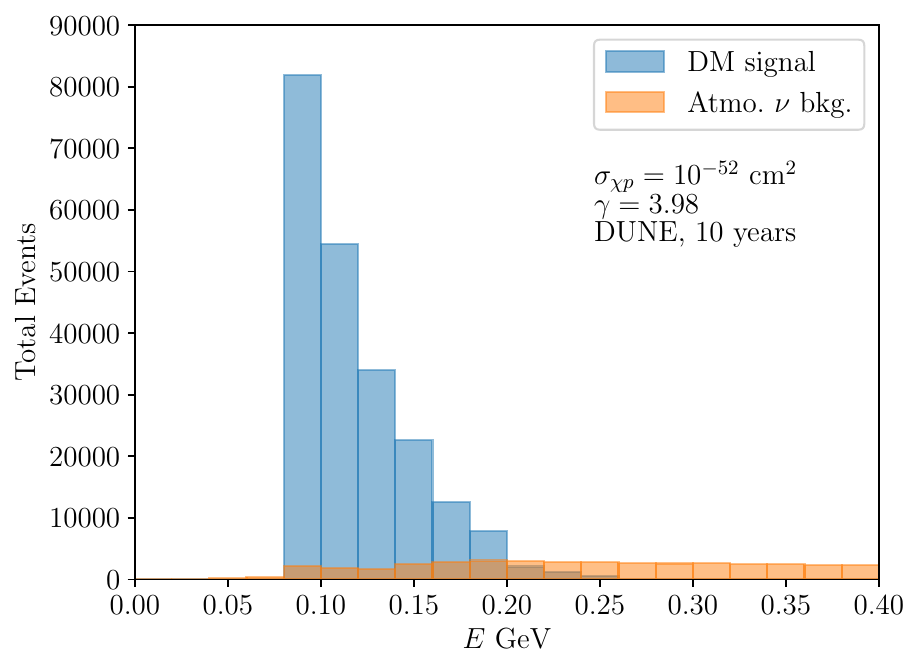}
\includegraphics[width=0.45\textwidth]{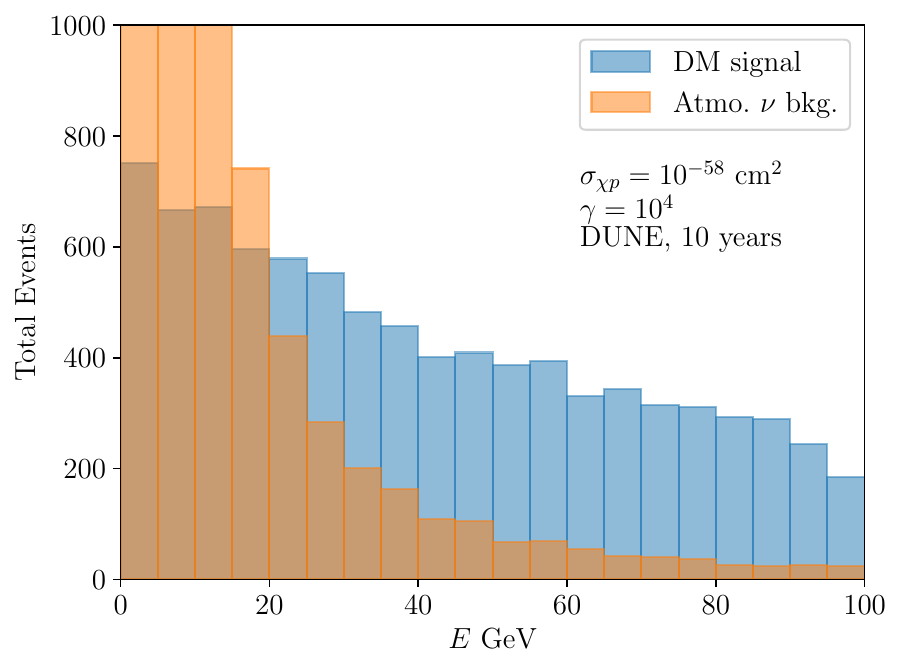}
\caption{Distribution of the total reconstructable energy at a boost of $\gamma = 3.98$ (left) and $\gamma = 10^4$ (right) and a DM mass of $100~\text{MeV}$.}\label{fig:energy}
\end{figure}

Given the expected interaction rate, we simulate boosted DM interactions in a detector using the \texttt{GENIE} software suite~\cite{Andreopoulos:2009zz,Andreopoulos:2015wxa,Berger:2018urf}. For our current study, we consider LZ, Super- and Hyper-Kamiokande, DUNE, JUNO, and IceCube/DeepCore sensitivities. These are the experiments likely to have the strongest sensitivity at the moment and in the coming years.  For each experiment, we consider benchmarks with DM masses of $1~\text{MeV}$, $10~\text{MeV}$, $100~\text{MeV}$, and $1~\text{GeV}$.  The \texttt{GENIE} simulation includes elastic, deep inelastic, and electron scattering (with resonant scattering a forthcoming feature). Given the simulated events, we select all visible particles above the thresholds indicated in Table \ref{tab:thresholds}. For LZ, the threshold is negligible compared to the energies we consider.  For IceCube and DeepCore, we are in the highly boosted regime and instead opt to apply a cut on the incident {\it dark matter} energy of 100 GeV and 10 GeV respectively. At these energies, deep inelastic scattering is very similar for neutrinos and for DM, so we apply the estimated neutrino energy thresholds to the DM. For the other detectors, we add up the momenta of these visible particles and reconstruct the total visible energy $E$. For protons, we only include the kinetic energy as the rest energy is not observed.  We calculate the angle $\theta$ of the total momentum with respect to the vertical and total energy. A distribution of angles and total energies for boosts of $\gamma = 3.98$ and $\gamma = 10^4$ in the long-range vector scenario are shown in Figures~\ref{fig:angle} and \ref{fig:energy}. These boosts correspond to the lowest ones allowing for elastic scattering and the highest ones we consider respectively.

\begin{table}[!tbh]
\centering
\begin{tabular}{l|l  l l l l}
\hline
Experiment & $\mu^\pm$ (MeV) & $\pi^\pm$ (MeV) & $p$ (MeV) & $e^\pm$ (MeV) & $\gamma$ (MeV) \\
\hline
DUNE & 35 & 35 & 80 & 30 & 30 \\
Super-K/Hyper-K & 55 & 75 & 485 & 3 & 3 \\
JUNO & 0.5 &  0.5 & 0.5 & 0.5 & 0.5 \\
\hline
\end{tabular}
\caption{Kinetic energy thresholds applied for charged particles and photons at the experiments studied in this work.  The thresholds for LZ are taken to be zero for all visible particles, as the actual thresholds are much lower than the particles produced in the DM interactions for the parameters considered. For IceCube and DeepCore, we instead place a threshold energy cut on of 100 GeV and 10 GeV on the {\it dark matter}, as described in the text.}\label{tab:thresholds}
\end{table}

In order to develop a meaningful cut on this sensitive variable, we also apply the same procedure to 500,000 atmospheric neutrino events below $E = 10~\text{GeV}$ and 500,000 atmospheric neutrino events between $E = 10~\text{GeV}$ and $E = 1000~\text{GeV}$. The background events are simulated using the Bartol flux calculation~\cite{Barr:2004br}. We include the high energy flux determined for Kamioka at all locations, as the flux varies little across the Earth at high energies. For the low energy flux, we use the Soudan flux for experiments at SURF and the Kamioka flux for experiments in Asia. Only the high energy fluxes, which are universal, are relevant for IceCube and DeepCore.
From these simulations, we are able to calculate the number of expected signal and background events for a livetime of 10 years at upcoming experiments and the current exposure of ongoing experiments.  We determine the cut on the angle with respect to the vertical and total energy that require the fewest signal events to satisfy the following conditions
\begin{equation}
    \frac{S}{\sqrt{B + \sigma_B^2 \, B^2}} \geq 2~,
\end{equation}
where $S$ is the total number of expected signal events for a given model point, $B$ is the total number of expected background events, and $\sigma_B$ is a relative background normalization uncertainty, which we take to be $0.3$ to account for systematic uncertainties. 
We take the best sensitivity of two possible cut flows, namely a) $\cos\theta > 0.8$, $E > 20~\text{MeV}$ and b) $\cos\theta > 0.9$, $E > 10~\text{GeV}$, with the total angle $\theta$ and energy $E$ as defined above. 

\begin{figure}
\centering
\includegraphics[width=0.49\textwidth]{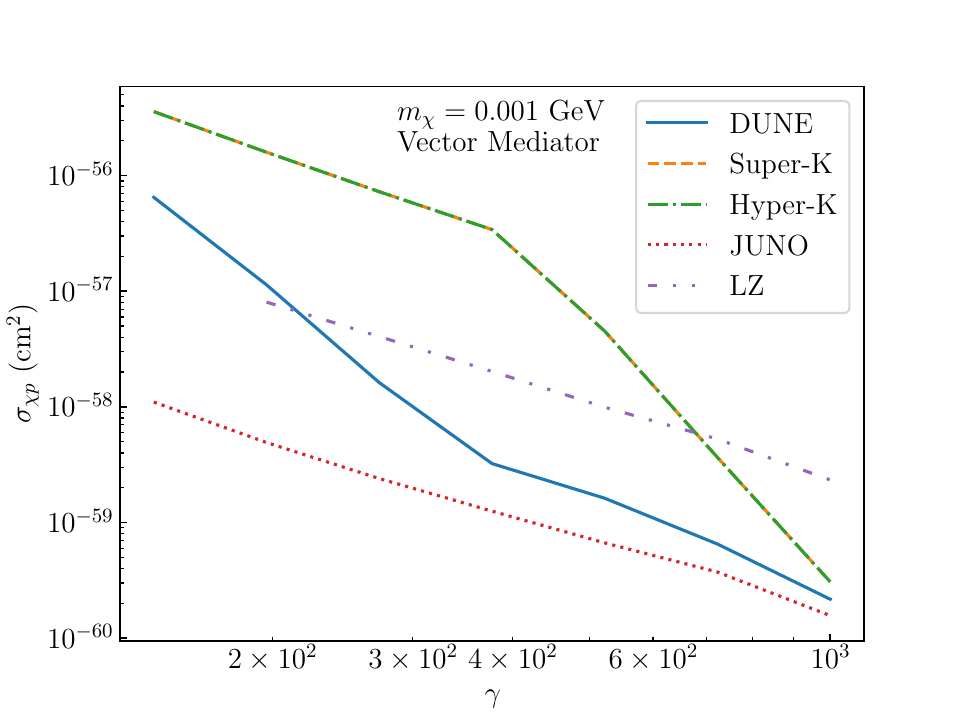}
\includegraphics[width=0.49\textwidth]{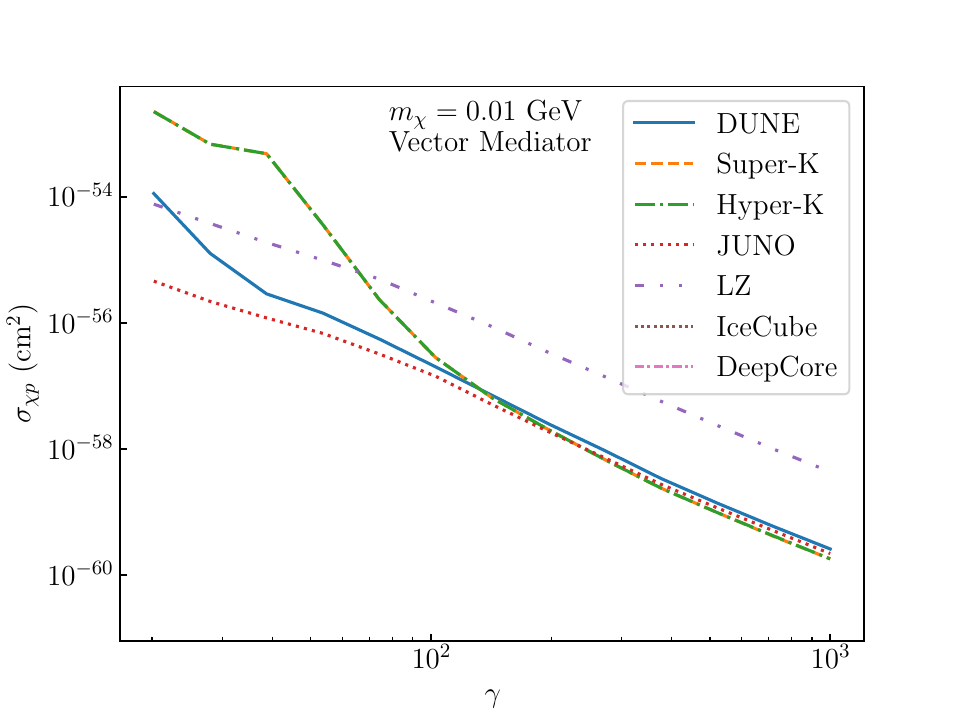}
\includegraphics[width=0.49\textwidth]{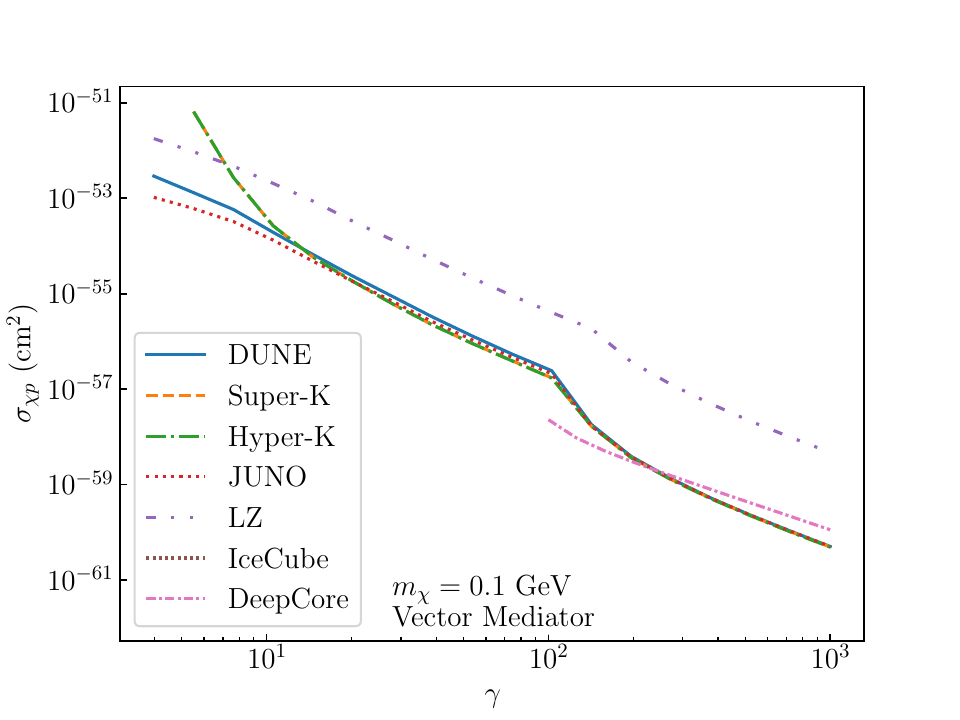}
\includegraphics[width=0.49\textwidth]{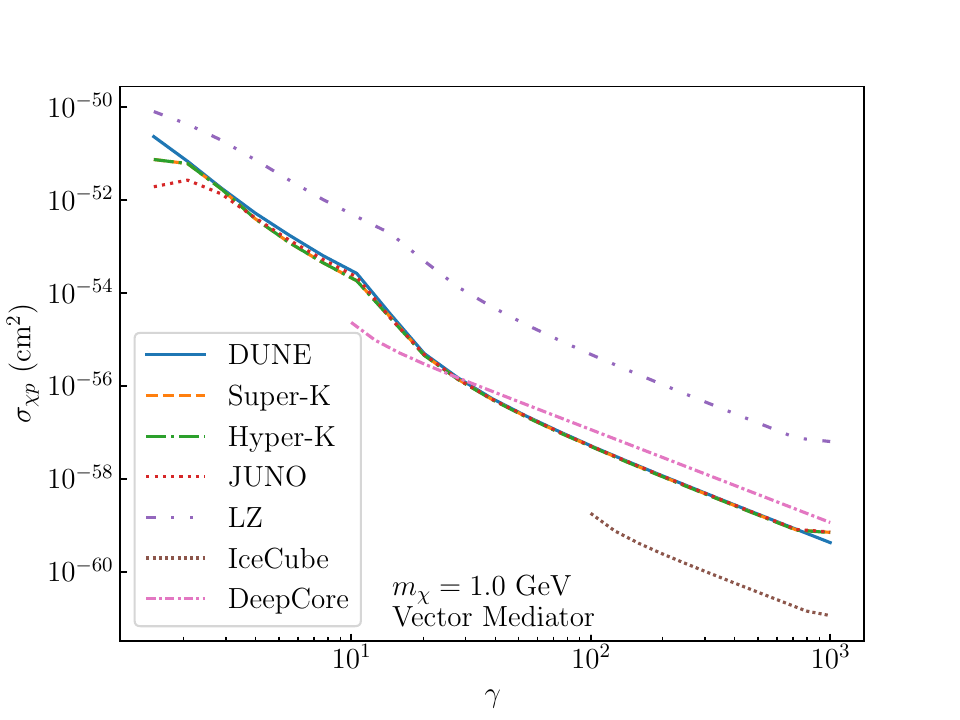}
\caption{The final sensitivity to the short-range interaction cross section as a function of the boost at the Earth's surface $\gamma$ for our DM rain model at different large volume detectors.
The different panels show the constraint for different DM masses: $m_\chi=1,10,100,1000$ MeV in the top left, top right, bottom left, and bottom right panels, respectively. The long-range mediator is assumed to be a vector.}
\label{fig:vector-sensitivity}
\end{figure}

\begin{figure}
\centering
\includegraphics[width=0.49\textwidth]{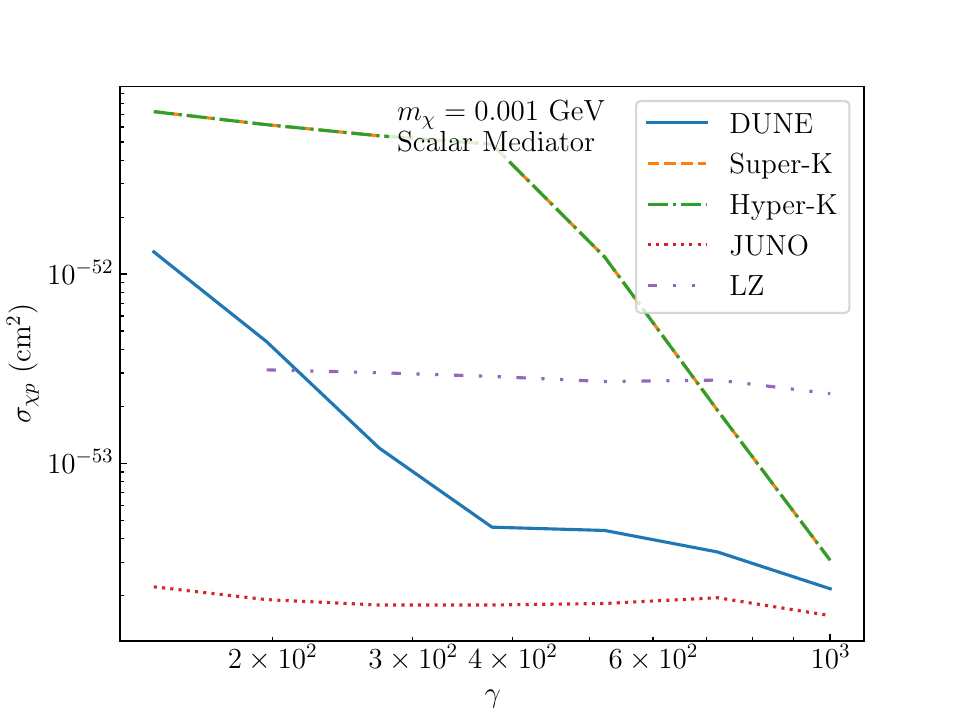}
\includegraphics[width=0.49\textwidth]{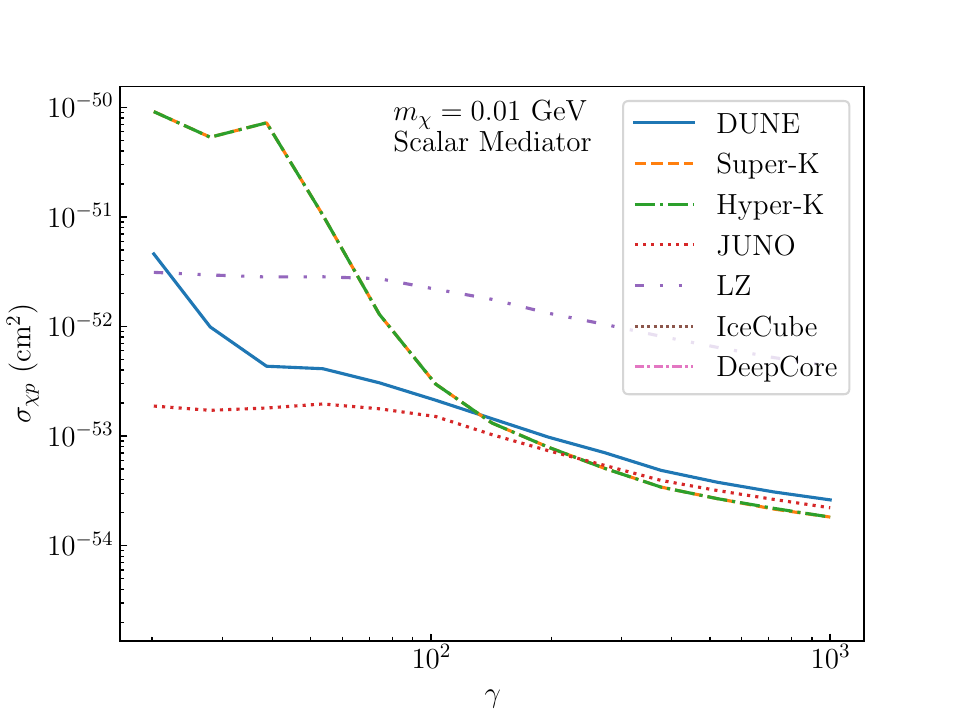}
\includegraphics[width=0.49\textwidth]{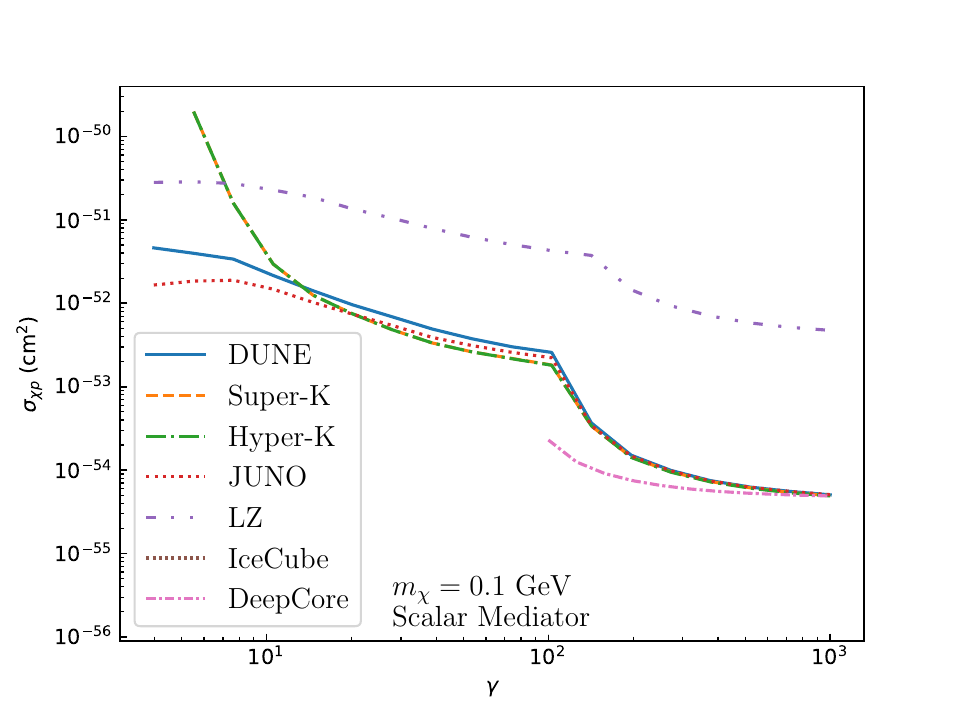}
\includegraphics[width=0.49\textwidth]{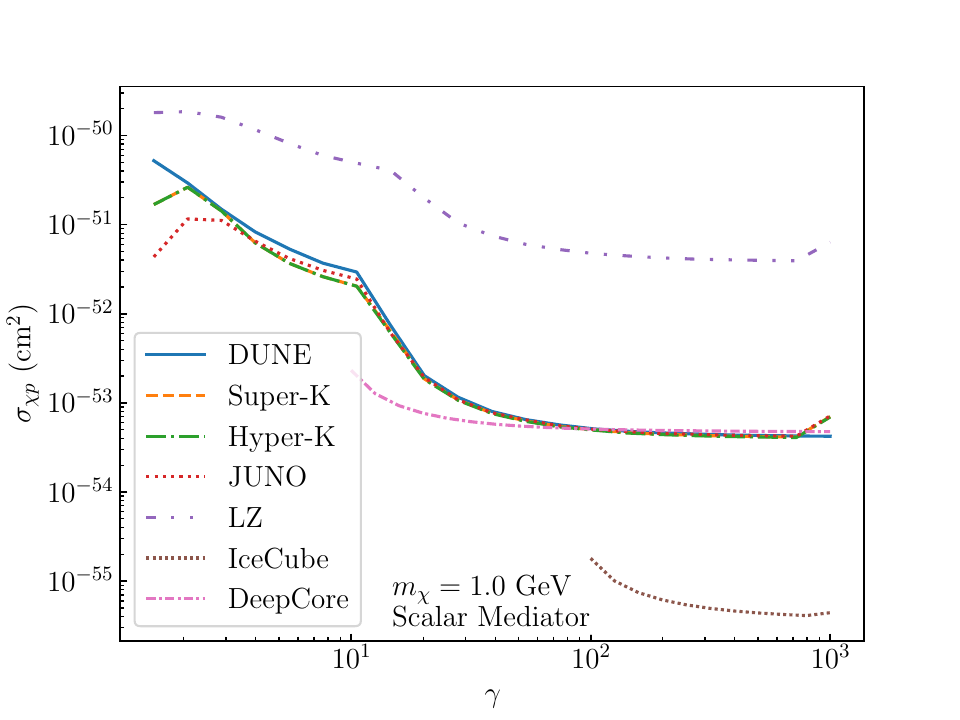}
\caption{The same as Fig.~\ref{fig:scalar-sensitivity} but for a scalar long-range mediator.}
\label{fig:scalar-sensitivity}
\end{figure}

We show our final sensitivity plots for various large volume detectors in Figures~\ref{fig:vector-sensitivity} and \ref{fig:scalar-sensitivity}.  In order to ease comparison between different experiments, we parameterize all models at a given boost using the cross section for interaction with protons $\sigma_{\chi p}$ at a given dark matter boost. This cross section is just a function of the short-distance interaction considered, rather than the specific nuclei in a given detector, and is assumed to be isospin-invariant. The cross sections with nuclei will be larger, though there will not be a coherent enhancement within the model we consider.  Roughly, it scales like the mass of the nucleus, though nuclear effects distort this scaling. As we emphasize the large boost regime, we have not determined the coherent scattering cross section, though this will be relevant at lower boosts.  The low boost regime motivates a careful determination of the threshold for coherent scattering at experiments like JUNO and DUNE that could be sensitive, in addition to consideration of higher energy recoils at direct detection experiments such as LZ. To match the normalization used in direct detection experiments, and to allow for fair comparison between experiments, we show our results in terms of the cross section per nucleon, rather than per nucleus. In each panel, we show the expected reach in terms of the short-range interaction as a function of the boost from the long-range interaction.

We generally see improved sensitivity at higher boosts due to the lower backgrounds and, in the case of vector mediators, enhanced flux.  Where the sensitivity curves overlap, we are limited by systematics, which may be able to be mitigated further beyond what we consider.  For example one could use the directions perpendicular to the vertical as a sideband to normalize the background.  At low boosts, we see a decrease in sensitivity in Water Cherenkov experiments due to the physical threshold of emitting Cherenkov light.  For LZ, which uses heavy Xenon nuclei as a target, elastic scattering turns off at low boosts.  For DeepCore and IceCube, we only draw sensitivity curves above the thresholds described above.  

Unsurprisingly, our DM rain model probes DM cross sections with the SM much smaller than traditional DM models.
A recent review found that constraints on sub-GeV DM without the additional long-range force are constrained at the $10^{-31}$ to $10^{-39}$ cm$^2$ level for DM with masses of 10 to 1000 MeV scattering off nucleons \cite{Essig:2022dfa}.
This can be qualitatively compared to our sensitivities at the $10^{-53}$ to $10^{-60}$ cm$^2$ level -- an increase of some 10 to 20 orders of magnitude -- depending on the underlying physics of the long-range force. 

Finally, we comment on additional astrophysical and cosmological considerations. Long-range forces have been considered in the context of dark kinetic heating of celestial bodies \cite{Gresham:2022biw,Acevedo:2024zkg}, resulting in constraints on DM-SM and DM-DM couplings. However, these limits require cross-section values much larger than what we probe here with large scale detectors, in order for a sizable fraction of the incoming DM to deposit energy and produce a observable temperature increase. For instance, the DM-nucleon cross-section at which a DM particle would scatter once while crossing a neutron star, is of the order of $10^{-45} \ \rm cm^2$, indeed several orders of magnitude above our lowest projections. The small cross-sections, however, may also imply that the DM was not thermalized with the SM bath in the early universe. Constructing a complete theory accounting for this would be beyond the scope of this work, but we note various works have developed models that admit such small cross-sections \cite{Evans:2019vxr,Belanger:2020npe,Compagnin:2022elr}.

\section{Conclusions}
\label{sec:outlook}
A number of highly instrumented large volume detectors are currently under construction around the world, such as JUNO, DUNE, and Hyper-K, as well as currently operating large detectors such as Super-K, IceCube and LUX-ZEPLIN which, with the exception of LZ, are predominantly focused on neutrino physics. Nonetheless, they can also contribute to world leading bounds on DM in certain scenarios, specifically some boosted DM scenarios.

In this paper, we have presented a detailed discussion of an interesting variation on boosted DM, dubbed DM rain. In this scenario, a dark sector long-range force accelerates halo DM towards massive celestial objects, such as the Earth. Unlike most other boosted DM scenarios, this causes 100\% of the DM at the Earth's surface to be highly boosted, with some Lorentz boost factor $\gamma$ which can be easily be $\mathcal O(100-1000)$ or larger. Because of the focusing effect of such a force, the DM is predominantly down-going and up-going, hence the name DM rain to indicate this unique phenomenological signature. 

We have detailed the phenomenological differences between a vector and a scalar long-range force, and how these impact on detection prospects of boosted DM on Earth. The vector long-range force leads to an increased flux relative to the scalar case for the same coupling strength, particularly in the high boost regime. We have also analyzed the unusual phenomenological impacts of a scalar long-range force when the effective DM mass induced by the interaction approaches zero. This occurs within a narrow region of parameter space, in terms of mediator mass and coupling, where notably large boost factors in the scalar case are achieved before additional radiative effects regularize the mass term, which we have not analyzed in detail. The DM radiative energy loss in this regime may lead to interesting phenomenology which is left for future investigation.

In addition to calculating the boost factor and flux at the Earth's surface, we have calculated the expected signal in various large volume detectors, using a modified \texttt{GENIE} code specifically suited to handle boosted DM-SM interactions. In particular, we find neutrino detectors to be extremely sensitive to such a scenario, particularly in the sub-GeV mass regime where they can even outperform DM experiments such as LZ. We have simulated boosts factors up to $10^3$, which encompass most of the open parameter space, and found DM-proton cross-section sensitivities extending down to $\sim 10^{-60} \ \rm cm^2$ can be achieved depending on the experiment and the specific long-range force model.
These sensitivities are tens of orders of magnitude stronger than traditional sub-GeV DM direct detection searches.

Finally, we have limited ourselves to force ranges larger than the Earth's radius, but smaller than about $1 \ \rm AU$, so that the anisotropic DM flux is aligned with the Earth and can be easily analyzed. Since this predicts a similar flux coming from above and below, it is unlikely to be confused by other backgrounds that are typically mostly downgoing. At larger force ranges, the interactions between the different celestial bodies in the Solar System may lead to more complex time-dependent anisotropies in the flux. Regardless, experiments can look for this distinctive signature separately from other boosted DM models considered in the past.

\acknowledgments
We thank Christopher Cappiello, Aidan Reilly and Erwin Tanin for helpful discussions. JFA is supported in part by the United States Department of Energy under Contract No.~DE-AC02-76SF00515.
PBD acknowledges support by the United States Department of Energy under Grant Contract No.~DE-SC0012704.

\appendix

\section{Radiation Losses}
\label{app:rad_thresh}
As DM is accelerated by the potential sourced by the Earth, it may radiate the light field carrier, leading to energy losses and therefore a reduced boost factor at detection. We show here that even for DM couplings at which large detectable boosts are attained, radiation losses are negligible. Specifically, we will estimate the energy lost to radiation for a given trajectory following the same procedure as for Larmor radiation (see $e.g.$ Ref.~\cite{Zangwill:1507229}). We will then compute the range of final boosts for which this lost energy is much smaller than the DM's kinetic energy gain. For simplicity, we will also assume the field to be massless; including a mass should further suppress the radiation rate, so our estimates below will be conservative. 

The instantaneous rate of radiated 4-momentum in an arbitrary frame is 
\begin{equation}
    \frac{dp_{\rm rad}^\mu}{d\tau} = - \mathcal{Q} \, a^\lambda a_\lambda \, U^\mu~,
\end{equation}
where the constant above is $\mathcal{Q}=g^2_{\chi}/6 \pi$ and $\mathcal{Q}=g^2_{\chi}/12\pi$ for a vector and scalar mediator, respectively, and $a^\lambda = dU^\lambda/d\tau$ is the DM's 4-acceleration. We can evaluate the $\mu = 0$ component in any frame, since the radiated power is a Lorentz-invariant. In particular, if we consider the Earth's rest frame then $U^{0} = \gamma$ and $d\tau = dt / \gamma$. This gives
\begin{equation}
    \frac{dE_{\rm rad}}{dt} = - \mathcal{Q} \, a^{\lambda} a_{\lambda}~,
\end{equation}
which is of course the covariant generalization of Larmor's formula. The contracted acceleration can be evaluated using the equations of motion, $cf.$ Eq.~\eqref{eq:eom_main}, which gives
\begin{equation}
    \frac{dE_{\rm rad}}{dt} = \mathcal{Q} \times \begin{cases}
      \  \dfrac{\left(U^\mu \partial^\lambda V_\mu - U^\mu \partial_\mu V^\lambda\right) \left(U^\mu \partial_\lambda V_\mu - U^\mu \partial_\mu V_\lambda\right)}{m_\chi^2} & \ \text{(vector)} \\
      \\
      \ \dfrac{\left(\partial^{\lambda} \Phi - \left(U^\mu\partial_{\mu}\Phi\right)U^\lambda\right)\left(\partial_{\lambda} \Phi - \left(U^\mu\partial_{\mu}\Phi\right)U_\lambda\right)}{(m_\chi + \Phi)^2} & \ \text{(scalar)} \\
      \end{cases}
\end{equation}
These expressions can be further simplified since the potential is constant, so all derivatives with respect to coordinate time vanish, and central, so the spatial gradient reduces to $\mathbf{\nabla} = (\partial/\partial{r}) \, \mathbf{\hat{r}}$ where $\mathbf{\hat{r}}$ is the radial unit vector. This yields
\begin{equation}
    a^\lambda a_\lambda = \begin{cases}
      \ - m^{-2}_\chi \, \gamma^2 \, \left(\cfrac{\partial V_0}{\partial r}\right)^2 \left(1 - v_\chi^2 \cos^2\theta \right) & \ \text{(vector)} \\
      \\
      \ - \left(m_\chi+\Phi\right)^{-2} \, \left(\cfrac{\partial \Phi}{\partial r}\right)^2 \, \left(1+\gamma^2 v_\chi^2 \cos^2\theta \right) & \ \text{(scalar)} \\
      \end{cases}
\end{equation}
where $\theta$ is the angle between the field gradient and the velocity 3-vector. This can be written in terms of the angular momentum using Eq.~\eqref{eq:cos_theta}, 
\begin{equation} 
\label{eq:app_cos_theta}
    \cos^2\theta = 1 - \left(\frac{b \, u}{r \, v_\chi \, \gamma}\right)^2~.
\end{equation}
Note that so far, $\gamma$ is the boost at some arbitrary point in the trajectory, which will be integrated to obtain the energy loss. In fact, for the calculations here, we will use the index $\oplus$ to denote the local boost and speed at the Earth's surface. In the vector scenario, energy loss maximally increases with boost when the particle moves perpendicular to the field gradient (see $e.g.$ Ref.~\cite{landau2013classical}). By contrast, in the scalar scenario energy loss with boost is maximal when the particle moves along the direction of the field gradient. To be conservative in our estimates, in the vector case we will assume the DM to have the maximum possible angular momentum at which it reaches a detector, while for the scalar case we will assume zero angular momentum. The latter case simply corresponds to fixing $\cos^2\theta = 0$. The former case corresponds to fixing $b = b_{\rm max} = R_\oplus \gamma_\oplus v^\oplus_\chi/u$ above, $cf.$ Eq.~\eqref{eq:max-impact}. For what follows, it is also useful to relate the coordinate $r$ to the boost using Eq.~\eqref{eq:eom_E} combined with Eqs.~\eqref{eq:potential_vec} and \eqref{eq:potential_sca}. Under the conservative assumption that the field is massless, this yields 
\begin{equation}
    r = \frac{\alpha N_{\oplus}}{m_\chi (\gamma - 1)} \times \begin{cases}
      \ 1 & \ \text{(vector)} \\
      \\
      \ \gamma & \ \text{(scalar)} \\
      \end{cases}
\end{equation}
Above, we have abbreviated $\alpha = g_{\rm SM} \, g_\chi/4\pi$. In particular, inserting the above in Eq.~\eqref{eq:app_cos_theta} for the vector interaction yields 
\begin{equation}
    \cos^2\theta = 1 - \left(\frac{R_\oplus v^\oplus_\chi \gamma_\oplus m_\chi (\gamma - 1)}{\alpha N_\oplus \gamma}\right)^2 \simeq 1 - \frac{1}{v_\chi^2} \left(\frac{\gamma - 1}{\gamma}\right)^2 ~,
\end{equation}
where, to obtain the rightmost side, we have approximated $v^\oplus_\chi \simeq 1$ and \begin{equation}
    \gamma_{\oplus} = 1 - \frac{V_0}{m_\chi} \simeq \frac{g_{\rm SM} g_{\chi} N_\oplus}{4 \pi m_\chi R_{\oplus}}~,
\end{equation}
which is already an excellent approximation at moderate boosts for the vector interaction ($cf.$ Eq.~\eqref{eq:boost_main}). 

We can now estimate the energy loss by integrating the radiated power $P$ over the unperturbed trajectory. Under the above assumptions it reads, 
\begin{equation}
    \label{app:Eloss_gen}
    \Delta E_{\rm rad} \simeq \mathcal{Q} \times \begin{cases}
      \ \displaystyle\int_{\infty}^{R_\oplus} \frac{\left(1 + (\gamma - 1)^2\right)}{m_\chi^2} \left(\frac{\partial V_0}{\partial r}\right)^2 \frac{dr}{v_\chi} & \ \text{(vector)} \\
      \\
      \ \displaystyle\int_{\infty}^{R_\oplus} \frac{\gamma^2}{\left(m_\chi+\Phi\right)^2} \left(\frac{\partial \Phi}{\partial r}\right)^2 \, \frac{dr}{v_\chi} & \ \text{(scalar)} \\
      \end{cases}
\end{equation}
It is convenient to recast the integral solely in term of the boost factor, which is then integrated from $\sim 1$ to the final boost reached at the Earth's surface. For both interaction types, we obtain
\begin{equation}
    \label{app:Eloss_gam}
    \Delta E_{\rm rad} = \mathcal{Q} \, \frac{m_\chi}{\alpha N_\oplus} \times \begin{cases}
      \ \displaystyle \int_{1}^{\gamma_\oplus} \frac{\gamma \left(1 + (\gamma - 1)^2\right) (\gamma - 1)^{3/2}}{\sqrt{\gamma + 1}} \, d\gamma & \ \text{(vector)} \\
      \\
      \ \displaystyle \int_{1}^{\gamma_\oplus} \frac{\gamma (\gamma - 1)^{3/2}}{\sqrt{\gamma + 1}} \, d\gamma & \ \text{(scalar)} \\
      \end{cases}
\end{equation}
For moderate boosts $\gamma_{\oplus} \gtrsim 5$, we can approximate the above integrands simply by $\gamma^n$, where $n = 4$ and $n = 2$ for the vector and scalar interactions respectively. Note that $\gamma_\oplus$ above is implicitly an increasing function of the effective coupling $\alpha N_\oplus$, so Eq.~\eqref{app:Eloss_gam} does not decrease with coupling as the prefactor $1/\alpha N_\oplus$ seems to imply.

In the long-range vector scenario, we compare the kinetic energy gain, of order $m_\chi \gamma_\oplus$ to the radiation losses. For the latter to be significant, it must be $\Delta E_{\rm rad} \simeq m_\chi \gamma_\oplus$. For moderate boosts, where the above integral can be well approximated as $\gamma_\oplus^5/5$, this condition can be expressed as 
\begin{equation}
    g^5_\chi \, g^3_{\rm SM} \simeq \frac{1920 \pi^4 (m_\chi R_{\oplus})^4}{N^3_\oplus}~.
    \label{eq:app-vec-rad-cond}
\end{equation}
The right hand side of Eq.~\eqref{eq:app-vec-rad-cond} is of order $\sim 10^{-72} \, (m_\chi/\rm MeV)^4$. By comparison, even maximizing the coupling product on the left hand side to current limits ($cf.$ Sec.~\ref{sec:BDM}), this is at most $\sim 10^{-90} \, (m_\chi/\rm MeV)^{15/4}$ assuming an electrophilic coupling. This indicates that $\Delta E_{\rm rad}$ cannot be a sizable fraction of the kinetic energy gain in our parameter space, and therefore radiative losses may be safely neglected. 

The scalar scenario admits an analysis similar to the above, although now we must compare the radiative energy loss to the mass defect $m_\chi - (m_\chi + \Phi(R_\oplus))$, which is order $m_\chi$ for the boosts we consider. This leads to a somewhat more complicated condition that must be met for radiation losses to be significant, 
\begin{equation}
    \frac{g_\chi}{9 \, g_{\rm SM} N_{\oplus}} \left(1 - \frac{g_{\rm SM} g_\chi N_\oplus}{4 \pi m_\chi R_\oplus}\right)^{-3} \simeq 1~,
    \label{eq:app-sca-rad-cond}
\end{equation}
where now, to derive this condition, we have approximated the integral in Eq.~\eqref{app:Eloss_gam} as $\gamma^3_\oplus/3$, and written the boost as 
\begin{equation}
     \gamma_{\oplus} = \frac{m_\chi}{m_\chi + \Phi} = \left(1 - \frac{g_{\rm SM} g_\chi N_\oplus}{4 \pi m_\chi R_\oplus}\right)^{-1}~.
\end{equation}
Unlike the vector scenario, the boost here cannot be approximated in a simple way because, for the largest values we have simulated of order $\sim 10^3$, the corresponding geometric series must be truncated at tens of thousands of terms for it to be accurate. Instead, we numerically verified that Eq.~\eqref{eq:app-sca-rad-cond} is not met by in the entire range of boosts we have simulated (we recall from Fig.~\ref{fig:boost} that maximizing the couplings will lead to cases beyond the critical point where the boost seems to diverge). In fact, for the largest boost we do consider, there is a relative factor of order $\lesssim 10^{-27}$ between both sides of Eq.~\eqref{eq:app-sca-rad-cond}, indicating that also in this scenario radiation losses are not relevant to our analysis.

\section{Centrifugal Barriers for Finite-Range Potentials}
\label{sec:centrifugal}
Due to the finite-range nature of the Yukawa potential, we must consider the possibility of multiple centrifugal barriers that may prevent DM particles with certain angular momenta from reaching the detector. This occurs when the contribution of the angular momentum in the effective potential becomes co-dominant with the Yukawa potential, as the latter begins to decay exponentially, inducing a local maximum. Such a maximum acts as a barrier if its value exceeds the kinetic energy of the incoming particle. In Eq.~\eqref{eq:max-impact}, we computed the maximum angular momentum, or equivalently, the impact parameter, for which a DM particle can reach the surface of the Earth without taking this effect into consideration. We now discuss this effect in detail and show that, for the typical parameters we consider, the DM always overcomes this secondary barrier. 

One starting point for analyzing this effect is the energy of the dark matter as a function of its radial speed. Using Eq.~\eqref{eq:HJ-2}, we can write the DM's energy as 
\begin{equation}\label{eq:app-E}
    E = \begin{cases}
      \  V_0 + m_\chi \, \sqrt{1 + \left(\dfrac{dr}{d\tau}\right)^2 + \left(\dfrac{L}{m_\chi r}\right)^2} & \ \text{(vector)}  \\
      \\
      \ \left(m_\chi+\Phi\right)\sqrt{1 + \left(\dfrac{dr}{d\tau}\right)^2 + \left(\dfrac{L}{\left(m_\chi+\Phi\right) r}\right)^2} & \ \text{(scalar)}  \\
    \end{cases}
\end{equation}
On the other hand, since $E$ is a constant of motion, we can equate it to the DM is non-relativistic far from the Earth, we have
\begin{equation}
    E \simeq m_\chi + \frac{1}{2} \, m_\chi \, u^2~.
\end{equation}

For a given angular momentum $L$, the turning point where the radial speed vanishes can then be solved by equating the above and setting $dr/d\tau = 0$. This can be done numerically, but under a few assumptions, we can obtain an approximate analytic solution. The main assumption we make is that the barrier peak occurs at $r \gg m^{-1}_{\phi, A^{\prime}} \,$. Numerically, we find this to be true, but this is also reasonable as the potential in either case must sufficiently decay for the centrifugal term to start dominating. Since the barrier peak also occurs far from the interaction range, we can also expand Eq.~\eqref{eq:app-E} in the non-relativistic limit, where both cases reduce to the usual Hamiltonian for a particle in a central potential. For the remainder of this analysis, we will then write the potential for either benchmark interaction as
\begin{equation}
    \Psi(r) = - \frac{\alpha \, N_\oplus}{r} \, e^{-m_{\phi, A^\prime} \, r},
\end{equation}
where $\alpha = g_{\rm SM} g_{\chi}/4\pi$. The total energy is given by 
\begin{equation}
   \frac{1}{2} \, m_\chi \left(\frac{dr}{dt}\right)^2 + V_{\rm eff}(r) = \frac{1}{2} \, m_\chi \, u^2~,
\end{equation}
where 
\begin{equation}
    V_{\rm eff}(r) = \frac{L^2}{2 \, m_\chi \, r^2} + \Psi(r)~
\end{equation}
is the effective potential. We can find the location of the centrifugal barrier by solving for the extrema, that is
\begin{equation}
   \frac{dV_{\rm eff}}{dr} \simeq \frac{L^2}{m_\chi \, r^3} + \frac{\alpha \, m_{\phi, A^\prime}}{r} \, e^{-m_{\phi, A^\prime} \, r} =  0~,
\end{equation}
where we have neglected the higher order term $\left(m_{\phi, A^\prime} \, r\right)^{-1}$. For the range of parameters we analyze, we find $m_{\phi, A^\prime} \, r \lesssim 0.1$ evaluated at the approximate position of the potential maximum. Under this approximation, depending on the model parameters, there are potentially two real positive solutions for the extrema of the effective potential at
\begin{equation}
    r_{\text{turn}} \simeq -\frac{2}{m_{\phi, A^\prime}} \times \begin{cases}
      \  W_{0}\left(-\beta\right) &  \\
      \\
      \ W_{-1}\left(-\beta\right) &  \\
    \end{cases}
\end{equation}
where $W_{k}$ is the $k^{\rm th}$ branch of Lambert's function, also known as product-log, and the argument $\beta$ is 
\begin{equation}
    \beta = \frac{L}{2} \, \sqrt{\frac{m_{\phi, A^\prime}}{\alpha \, m_\chi}}~.
\end{equation}

The $0^{\rm th}$ branch of the product-log corresponds to the local minimum, while the $1^{\rm st}$ branch gives the maximum we are interested in. Both solutions are real and positive provided
\begin{equation}
     \beta < \frac{1}{e} \simeq 0.367~.
\end{equation}
The potential barrier at the local maximum is approximately
\begin{equation}
    V_{\rm eff}(r_{\rm turn}) \simeq \frac{L^2 \, m^2_{\phi, A^\prime}}{8 \, m_\chi} \, \left(\frac{W_{-1}(-\beta) + 1}{W^3_{-1}(-\beta)}\right)~.
\end{equation}

In terms of the impact parameter, the condition for a DM particle passing over this barrier is
\begin{equation}
    \frac{b \, m_{\phi, A^\prime} }{2} \lesssim \sqrt{\frac{W_{-1}^3(-\beta)}{W_{-1}(-\beta) + 1}}~,
    \label{eq:2nd_barrier_threshold}
\end{equation}
where we have used above Eq.~\eqref{eq:orb_ang_mom}. The speed dependence cancels out when demanding that the kinetic energy is sufficient to overcome the local potential maximum.

The maximum impact parameter $b_{\rm sec}$ for which Eq.~\eqref{eq:2nd_barrier_threshold} is satisfied is to be compared to Eq.~\eqref{eq:max-impact} in the main text. Consequently, the actual impact parameter for which a DM particle reaches the surface of the Earth is
\begin{equation}
    b_{\rm det} = {\rm min} \left[b_{\rm max}, b_{\rm sec}\right]~.
\end{equation}
The $\rm min$ function indicates that the impact parameter must be capped to the maximum value for which the DM particles are able to pass through the local potential maximum at $r = r_{\rm turn}$. Since the impact parameter determines the flux via the integration of Eq.~\eqref{eq:flux_main} over angular momentum, this effect can dominate the DM flux passing through a detector. Whether $b_{\rm det}$ saturates to $b_{\rm sec}$ will depend on the specific model parameters. This can be particularly the case for the vector interaction, where $b_{\rm max}$ is significantly enhanced relative to the scalar case by the boost factor at the surface. For boosts $\gtrsim 10^3$, mediator masses $m^{-1}_{A^\prime} \lesssim 1 \ \rm AU$ result in the impact parameter maxing out to $b_{\rm sec}$, limiting the flux to a value considerably lower than what is obtained from $b_{\rm max}$. While we could consider lighter mediators, for ranges beyond $\sim 1 \ \rm AU$ neighbouring celestial bodies may impact on the DM flux as noted in Sec.~\ref{sec:BDM}, which would further complicate calculations. For the scalar case, we find this effect to be negligible in the range of interest $R_\oplus \ll m^{-1}_\phi \lesssim 1 \ \rm AU$. Because of this, we do not consider the regime where the impact parameter $b_{\rm sec}$ dominates the flux, since it is either irrelevant or corresponds to boost factors beyond what we have simulated. For the sensitivity projections shown in Figs.~\ref{fig:vector-sensitivity} and \ref{fig:scalar-sensitivity}, it suffices to note that a wide range of mediator masses $R_{\oplus} \ll m^{-1}_{\phi,A^\prime} \lesssim 1 \ \rm AU$ is allowed where this effect does not dominate the flux. Moreover, even in the parameter space where this local potential maximum is indeed relevant, we still expect a considerable flux enhancement relative to the purely gravitational case, resulting in optimistic detection prospects. 

Finally, we also note that these analytic estimates done here might change if additional effects, such as field screening from quartic terms in Eqs.~\eqref{eq:mod_vec} and \eqref{eq:mod_sca}, are introduced. Whether these effects are important or not will depend on the parameter space and the specific model realization (see $e.g.$ Ref.~\cite{Acevedo:2024zkg}), so we do not explore these scenarios in detail. 

\section{Flux Calculation}
\label{sec:Gould_recap}
We provide additional details on how Eq.~\eqref{eq:flux_main} is derived, following the analysis of Ref.~\cite{Gould:1987ir}. First, we consider a sphere of radius $R$ much larger than the range of the long-range force, so that the incoming DM is non-relativistic. The differential flux of DM particles with speed $u$ to $u + du$ passing through this surface, at angle $\theta$ to $\theta + d\theta$ relative to the radial direction, is given by
\begin{equation}
\label{eq:app-flux}
d\mathcal{F} = \frac{1}{2} \, f(u) \, u \, \cos\theta \, d\!\left(\cos\theta\right)~.
\end{equation}
where $f(u)$ is the velocity distribution. Of course, only $0 < \cos\theta < 1$ should be considered, as negative $\cos\theta$ corresponds to leaving the sphere. This can be expressed in terms of the conserved angular momentum per unit rest mass $J$ evaluated at $r = R$,
\begin{equation}
J = R \, u \, \sin\theta~.
\end{equation}
Then, one can trade $\cos\theta$ for $J^2$ as the integration variable
\begin{equation}
\label{eq:app-dJ}
d\!\left(\cos\theta\right) = \frac{1}{2} \, d\!\left(\cos^2\theta\right) = - \frac{1}{2} \, \frac{dJ^2}{R^2 u^2}~.
\end{equation}
Plugging Eq.~\eqref{eq:app-dJ} into Eq.~\eqref{eq:app-flux}, and multiplying by the surface area of the sphere $4 \pi R^2$, we recover Eq.~\eqref{eq:flux_main} in the main text. The sign above cancels out upon inverting the integration limits with respect to $J$, and so we omit it.
            
\bibliographystyle{JHEP}
\bibliography{biblio}

\end{document}